\documentclass[structabstract]{aa}
\usepackage{times}
\usepackage{paralist}
\usepackage{float}
\usepackage{array}
\usepackage{tabularx}
\usepackage{changepage}
\usepackage{color}
\usepackage[latin1]{inputenc}
\usepackage{graphicx}
\usepackage{natbib}
\usepackage{mathabx}
\usepackage{multirow}
\usepackage{txfonts}
\usepackage{bm}

\def\degr{\hbox{$^\circ$}}
\def\arcmin{\hbox{$^\prime$}}
\def\arcsec{\hbox{$^{\prime\prime}$}}
\begin{document}

   \title{CHANG-ES IX:}

   \subtitle{Radio scale heights and scale lengths of a consistent sample of 13 spiral galaxies seen edge-on and their correlations}

   \author{Marita Krause\inst{1}, Judith Irwin\inst{2}, Theresa Wiegert\inst{2}, Arpad Miskolczi\inst{3}, Ancor Damas-Segovia\inst{1},
          Rainer Beck\inst{1}, Jiang-Tao Li\inst{4}, George Heald\inst{5}, Peter M\"uller\inst{1}, Yelena Stein\inst{3}, Richard J. Rand\inst{6}, Volker
          Heesen\inst{7}, Rene A. M. Walterbos\inst{8}, Ralf-J\"urgen Dettmar\inst{3}, Carlos J. Vargas\inst{8},
          Jayanne English\inst{9}
          \and
          Eric J. Murphy\inst{9}
          }

   \institute{Max-Planck-Institut f\"ur Radioastronomie, Auf dem H\"ugel 69, 53121 Bonn, Germany\\
              \email{mkrause@mpifr-bonn.mpg.de}
         \and
              Dept. of Physics, Engeneering Physics, \& Astronomy, Queen's University,
              Kingston, Ontario, Canada, K7L 3N6\\
              \email{irwin@astro.queensu.ca}
         \and
              Astronomisches Institut, Ruhr Universit\"at Bochum, 44780 Bochum, Germany
         \and
              Department of Astronomy, University of Michigan, 311 West Hall, 1085 S. University Ave, Ann Arbor, MI, 48109-1107, U.S.A.
         \and
              CSIRO Astronomy and Space Science, 26 Dick Perry Avenue, Kensington, WA 6151, Australia
         \and
              Department of Physics and Astronomy, University of New Mexico, 800 Yale Boulevard, NE, Albuquerque, NM, 87131, U.S.A.
         \and
              Universit\"at Hamburg, Hamburger Sternwarte, Gojenbergsweg 112, 21029 Hamburg, Germany
         \and
              Department of Astronomy, New Mexico State University, PO Box 30001, MSC 4500, Las Cruces, NM, 88003, U.S.A.
         \and
              Department of Physics and Astronomy, University of Manitoba, Winnipeg, Manitoba, R3T 2N2, Canada
         \and
             Department of Astronomy, PO Box 400325, 530 McCormick Road, Charlottesville, VA 22904-4325
             }

   \date{Received ; accepted }


  \abstract
   {}
   {The vertical halo scale height is a crucial parameter to understand the transport of cosmic-ray electrons (CRE) and their energy loss mechanisms in spiral
    galaxies. Until now, the radio scale height could only be determined for a few edge-on galaxies because of missing sensitivity at high
    resolution.
   }
   {We developed a sophisticated method for the scale height determination of edge-on galaxies. With this we determined the scale
    heights and radial scale lengths for a sample of 13 galaxies from the CHANG-ES radio continuum survey in two frequency bands.
  }
   {The sample average value for the radio scale heights of the halo are $1.1\, \pm \, 0.3$~kpc in C-band and $1.4\, \pm \, 0.7$~kpc
    in L-band. From the frequency dependence analysis of the halo scale heights we found that the wind velocities (estimated using
    the adiabatic loss time) are above the escape velocity. We found that the halo scale heights increase linearly with the
    radio diameters. In order to exclude the diameter dependence, we defined a normalized scale height $\tilde{h}$ which is quite
    similar for all sample galaxies at both frequency bands and does not depend on the star formation rate or the magnetic field  
    strength. However, $\tilde{h}$ shows a tight anticorrelation with the mass surface density.
  }
   {The sample galaxies with smaller scale lengths are more spherical in the radio emission, while those with larger scale lengths are
    flatter. The radio scale height depends mainly on the radio diameter of the galaxy. The sample galaxies are consistent with an
    escape-dominated radio halo with convective cosmic ray propagation, indicating that galactic winds are a widespread phenomenon
    in spiral galaxies. While a higher star formation rate or star formation surface density does not lead to a higher wind velocity, we
   found for the first time observational evidence of a gravitational deceleration of CRE outflow, e.g. a lowering of the wind velocity
   from the galactic disk. 
   }


               \titlerunning{Radio scale heights in galaxy halos}

               \authorrunning{M. Krause et al.}
               \maketitle

\section{Introduction}
\label{sec:introduction}\titlerunning{Radio scale heights in the halo}

The halo around spiral galaxies is the interface between the bright disk with star formation activity and the intergalactic medium    
and can best be observed in edge-on galaxies.
Halos of star-forming galaxies are formed of stars, gas, cosmic rays, and magnetic fields. The star formation in the galactic disk is generally regarded to be the
source of cosmic-ray electrons (CRE) accelerated in the shock fronts of supernova (SN) explosions. Together with the magnetic field, CRE are responsible for 
non-thermal radio emission. The gravitational force is balanced by thermal, non-thermal, 
and kinetic pressures. Models assuming hydrostatic equilibrium are used to describe the global state 
of the multi-phase interstellar medium (ISM) in galactic halos \citep{fletcher+2001, boulares+1990}.
The assumption of hydrostatic equilibrium appears to be valid when considering the average state of the ISM over large scales or times.
Magnetic fields and cosmic rays are generally thought to be important contributors to the pressure balance of galaxy halos \citep{boulares+1990},
while the role of gas outflows and their kinetic pressure is unclear.

The distributions of gas density, cosmic rays, and magnetic fields with distance from the galaxy plane can be described by at least two exponential functions,
one with a small scale height (typically a few 100~pc) for the disk itself and one with a larger scale height (1--2~kpc) for the halo. Magnetohydrodynamic models of 
evolving galaxies indicate that another very extended component with a scale height of 10--40~kpc may exist \citep{pakmor+2017}.

The best tool for measuring the energy densities of the magnetic fields and cosmic rays is non-thermal (synchrotron) emission, which is strongest in the radio range.
The intensity of synchrotron emission is proportional to the magnetic field strength to the power of about two and the number density of CRE
in the relevant energy range.  CRE with energies of a few GeV emit radio waves in the GHz range in magnetic fields of about 10~$\mu$G strength.
The total radio continuum emission of a galaxy halo is dominated by this synchrotron emission.

The extent of a radio halo perpendicular to the disk of a spiral galaxy, as we observe it in the radio continuum, is related to
the extent of magnetic fields and CRE and to the sensitivity level of a radio map. Usually, the magnetic field extends far out into the halo and further into the 
intergalactic medium (IGM). Under the assumption of energy equipartition between cosmic rays and magnetic fields, the extent of magnetic fields is larger than that of 
cosmic rays by a factor of about 2 \citep{beck2007}.

The extent of a radio halo perpendicular to the disk of a spiral galaxy, as we observe it in the radio continuum, is certainly related
to the sensitivity level of the map. The radio emission of a halo is dominated by non-thermal synchrotron emission which
requires a magnetic field and CRE. Usually, the magnetic field extends much further out into the halo and further into the
IGM than the CRE (with the assumption of energy equipartition between CRE and the magnetic field), even by a factor of about 2 \citep{beck2007}.
In addition, the CRE lose their energy while they travel outwards into the halo and their volume density decreases due to adiabatic expansion into the IGM. Both effects
lead to a `halo boundary' which is frequency dependent. Beyond this boundary synchrotron emission is no longer detectable because CRE are not energetic enough or not
sufficiently numerous. Depending on which process dominates, we distinguish between a synchrotron energy loss-dominated (low energy) or an escape-dominated  halo (a few CRE).

The transport of the CRE themselves can happen by diffusion or by convective outflow. These processes have different
frequency dependences which can be used to identify them. They also depend on the total magnetic field strength and field
ordering \citep{taba+2013}.

From pioneering observations of edge-on galaxies by \cite{hummel+1989} and \cite{hummel+1991b,hummel+1991a} it is clear that radio continuum
emission can extend far into galaxy halos up to vertical heights of many kpc. As the simple extent of a radio halo depends on the sensitivity of the
observations, a better suited and more physical quantity is usually taken to describe the halo size. This is the radio scale height of the halo,
sometimes also referred to as scale height of the thick disk \citep{dumke+1995}.

Until now, the radio scale heights of the halos could only be determined for a few bright and extended nearby spiral galaxies (see \citealt{dumke+1998},
\citealt{krause2009}, \citealt{heesen+2016}). With the  Continuum HAlos in Nearby Galaxies -- an EVLA Survey (CHANG-ES) 
observations \citep{irwin+2012a},
we significantly increased the number of observed edge-on galaxies, as well as the sensitivity of the radio maps. In this paper we present
the scale heights of thirteen galaxies of this sample in C-band and L-band (with central frequencies of 6~GHz and 1.5~GHz) by which the number of observed edge-on 
galaxies increases significantly, especially towards smaller or more distant objects. We describe the sample selection in Sect.~\ref{sec:sample} and present the
method used for the scale height determination in Sect.~\ref{sec:method}. The results are presented in Sect.~\ref{sec:results} and compared with other structural
properties of the galaxies in Sect.~\ref{sec:scale length}. We also determined the magnetic field
strengths for the galaxies, given in Sect.~\ref{sec:field strength}, and discuss the frequency dependence of the halo scale heights in Sect.~\ref{sec:frequency}.
Finally, the radio scale heights were tested for correlations with the magnetic field strengths and star formation in Sect.~\ref{sec:correlations}
and discussed in Sect.~\ref{sec:discussion}, followed by the summary and conclusion in Sect.~\ref{sec:summary}.


\section{ Galaxy sample}
\label{sec:sample}
CHANG-ES \citep{irwin+2012a} is an unprecedented deep
radio continuum survey of a sample of 35 nearby spiral galaxies seen edge-on and observed with the Karl G. Jansky Very Large
Array (hereafter Expanded VLA or EVLA) in its commissioning phase. The sample galaxies are selected from the Nearby
Galaxies Catalog (NBGC; \citealt{tully1988}) according to their radio flux density at 1.4 GHz ($ S_{1.4}> 23$~mJy), blue isophotol
diameter ($4\arcmin < d_{25} < 15\arcmin$), declination ($\delta > -23\degr$), and inclination ($i > 75\degr$) (see
Table~1 in \citealt{irwin+2012a}). In addition, three well-known edge-on galaxies just outside the size limits were also
added: NGC~4244, NGC~4565, and NGC~5775. There are 35 galaxies in total which enable us, for the first
time, to observe radio halos and determine vertical scale heights in a statistically meaningful sample of nearby spiral
galaxies. The galaxies were observed in   C- and L-band at B- (only L-band), C-, and D-array configurations (see \cite{irwin+2012a}
for details). The total and polarized intensity maps at D-array are published in Paper IV \citep{wiegert+2015}
and are available in the CHANG-ES Data Release I \footnote {The CHANG-ES data release I is available at www.queensu.ca/changes}.
The observations at C-array are also reduced and will be published by Walterbos et al. (in preparation). As we need high
sensitivity and high resolution for the determination of the vertical radio scale heights, we analyse in this paper C-band
(6~GHz with 2~GHz bandwidth) D-array and L-band (1.5~GHz with 500~MHz bandwidth) C-array maps of total intensity with an
(untapered) angular resolution between 9\arcsec\ and 16\arcsec.

Missing large-scale flux density for extended structures larger than 4\arcmin~is expected at C-band. As we do not yet have corresponding
single-dish observations to combine with our CHANG-ES data, we excluded eight galaxies from our sample with angular sizes larger than 
$12\arcmin$ (as determined from the plots in Paper IV \citep{wiegert+2015}). We omitted five more galaxies which do not show extended 
disk emission and are known to have strong nuclear
activity and/or to be in a merger phase:  NGC~660, NGC~2992, NGC~4594, NGC~4845, and NGC~5084. The initial value for the
position angle of each galaxy was taken from NED and the literature. For each galaxy we checked these values with our new radio observations
at C-band.

As the scale heights are meant to describe the vertical extent of the disk and halo of a galaxy, we have to restrict
ourselves to strongly inclined objects. The value
for the inclination  given in \citet{tully1988} was checked for each galaxy with that given in NED
and actual values in the literature. As our scale height program (see Sect.~\ref{sec:method}) is very sensitive to
this parameter it even allows us to verify the value within a certain range. Ultimately, we found that our best estimates
for the inclination deviate up to 8\degr~from the values given by \cite{tully1988}. We also found
that vertical scale heights can only properly be fitted for galaxies with inclinations $i > 80\degr$ with the correction
for the disk emission applied in our model (see Sect.~\ref{sec:method}). Hence, we rejected seven galaxies from our sample
which we found to have inclinations $i < 80\degr$:
NGC~3448, NGC~4388, NGC~4096, NGC~4438, NGC~4666, NGC~5297, and NGC~5792. This is probably also the case for two more galaxies,
NGC~2613 and NGC~2683. In addition, the latter shows a varying position angle by $\sim 25 \degr$ across the velocity field, which
may indicate a strong bar viewed close to side-on \citep{kuzio+2009}. Hence, we also exclude these two galaxies. Our final galaxy
sample contains 13 edge-on spiral galaxies which are listed in Table~\ref{sample} together with our best estimates for the
inclination ($i$) and position angles ($p.a.$). The values for the distance are taken from Paper IV \citep{wiegert+2015}.
The coordinates given in Table~\ref{sample} are the central positions of the galaxies as given in \cite{wiegert+2015} except for
NGC~4013, for which we determined the location of the strong nuclear source from our $9.1 \arcsec$ resolution map to be
$\alpha_{2000}=11^h 58^m 31.^s4$ and $\delta_{2000}= 43\degr 56\arcmin 50\arcsec$.
All maps (at C- and L-band) were smoothed to a circular beam with $HPBW$ given in Table~\ref{sample} together with their rms noise
values in our maps. The rms values listed in Cols. 8 and 10 were measured from our slightly smoothed primary beam
corrected maps. They are thus different from the values given in \cite{wiegert+2015} for C-band and Walterbos et al. (in prep.) for
L-band, which were measured from the maps without primary beam correction and far from the source.
\\

\begin{table*}
      \caption[]{\label{sample}  Galaxy sample}
      $$
         \begin{tabular}{lccccccccc}
            \hline
            \noalign{\smallskip}
            \multicolumn{1}{c}{\textbf{Source}} &  &  &  &   &   &\multicolumn{2}{c}{\textbf{C-band D-array}}   &\multicolumn{2}{c}{\textbf{L-band C-array}} \\
                        & RA & Dec &    $i$   &    $p.a.$   & Distance$^\mathrm{a}$
                           & HPBW       & rms           & HPBW       &  rms          \\
                        &  [h m s]  & [\degr\  \arcmin\  \arcsec\ ] &  [\degr]  &  [\degr]  & [Mpc]  & [\arcsec]  & [$\mu$Jy/beam] & [\arcsec]  & [$\mu$
                        Jy/beam] \\
            \noalign{\smallskip}
            \hline
            \noalign{\smallskip}
            NGC~2820  &  09 21 45.6  &  64 15 29  &  88  &  65              & 26.5  & 10.0  &  9     & 10.0  &  25 \\
            NGC~3003  &  09 48 36.1  &  33 25 17  &  85  &  75              & 25.4  &  9.4  & 11     &  9.9  &  25 \\
            NGC~3044  &  09 53 40.9  &  01 34 47  &  85  & -67              & 20.3  & 10.7  & 10     & 11.5  &  30 \\
            NGC~3079  &  10 01 57.8  &  55 40 47  &  84  & -13              & 20.6  &  9.4  & 12     & 10.1  & 100 \\
            NGC~3432  &  10 52 31.1  &  36 37 08  &  85  &  41              &  9.4  & 12.0  & 12     & 10.2  &  30 \\
            NGC~3735  &  11 35 57.3  &  70 32 08  &  84  & -52              & 42.0  & 10.1  & 10     & 10.6  &  20 \\
            NGC~3877  &  11 46 07.8  &  47 29 41  &  85  &  35              & 17.7  &  9.0  & 11     &  9.9  &  22 \\
            NGC~4013  &  11 58 31.4  &  43 56 50  &  88  &  65              & 16.0  &  9.1  & 10     & 10.0  &  25 \\
            NGC~4157  &  12 11 04.4  &  50 29 05  &  83  &  66              & 15.6  &  9.2  & 12     & 10.1  &  35 \\
            NGC~4217  &  12 15 50.9  &  47 05 30  &  86  &  50              & 20.6  &  9.0  &  9     & 16.4  &  25 \\
            NGC~4302  &  12 21 42.5  &  14 35 54  &  90  &   0              & 19.4  & 10.0  & 25     & 10.5  &  30 \\
            NGC~5775  &  14 53 58.0  &  03 32 40  &  86  & -35              & 28.9  & 10.7  & 10     & 11.5  &  25 \\
            U10288    &  16 14 24.8  & -00 12 27  &  90  &  90              & 34.1  & 11.0  & 14     & 12.2  &  20 \\

            \noalign{\smallskip}
            \hline
          \end{tabular}
          $$
     
\begin{list}{}{}
\item[$^{\mathrm{a}}$] taken from Paper IV \citep{wiegert+2015}
\end{list}
   \end{table*}

\section{ Method}
\label{sec:method}

The determination of the vertical radio scale heights of spiral galaxies seen edge-on
goes back to the method initially described and applied by \cite{dumke+1995}. We determined the scale heights with the new NOD3 program
package \citep{mueller+2017}. The task is called `BoxModels' and works with total intensity maps in fits-format smoothed to a circular beam.

Within the task, the galaxy is rotated by an angle given by the position angle of the galaxy to a horizontally orientated major
axis. Then strips are chosen perpendicular to the major axis (hence in the z-direction) positioned with respect to the centre of the galaxy.
As an example we use our 6~GHz map of NGC~5775 (Fig.~\ref{n5775map}) with the strips on top of the rotated galaxy in Fig.~\ref{n5775map.rotated}.
Each strip consists of a number of rectangular boxes, specified by the width of the strip and a certain height
within which the radio intensities are averaged. The uncertainties in the intensities were derived from the standard deviation around the mean in
each box. In order to exclude the intensity gradient in the z-direction from the standard deviation, the standard deviation is calculated separately along
each row within a box parallel to the galaxy's major axis, and averaged over all rows within each box.
The averaged intensity in each box forms a distribution with z that is fitted to determine the scale height along each individual strip.
For inclinations $i < 90\degr$ the observed intensity distribution is the superposition of the intrinsic vertical distribution of the disk
and halo emission and the inclined disk emission (even if we  assumed that the disk is infinitesimally thin),
convolved with the beam of the telescope. As we are only interested in the vertical intensity distribution, the contribution of the
disk has to be estimated. It depends only on the disk's diameter and inclination.
The disk is assumed to be circular in shape with a diameter given
by its radio emission along the major axis. Its inclined values are used for a diameter-dependent
geometrical correction for the contribution of the disk on the scale heights by defining an effective beam size $HPBW_{\mathrm{eff}}$
which is largest at the central strip and decreases in the strips further outwards along the major axis approaching the value of the $HPBW$ at the edges.

As an exact deconvolution of the observed intensity distribution with $HPBW_{\mathrm{eff}}$ cannot be applied, we assume
different possible vertical distributions convolved with the corresponding $HPBW_{\mathrm{eff}}$ for each strip and compare them
with the observed distributions by least-square fits. Similar to \cite{dumke+1995}, we assume an intrinsic exponential

\begin{equation}
 w_{\mathrm{exp}}(z)=w_{0}\exp{(-z/h)}
\end{equation}

or an intrinsic Gaussian distribution

\begin{equation}
 w_{\mathrm{Gauss}}(z)=w_{0}\exp{(-z^{2}/h^{2})}
\end{equation}
with peak intensity $w_{0}$ and scale height $h$. This is convolved with the effective telescope beam
$(HPBW_{\mathrm{eff}} = 2\sqrt{2\ln{2}}\cdot\sigma_{\mathrm{eff}})$

\begin{equation}
\label{beam}
 g(z)=\frac{1}{\sqrt{2\pi\sigma_{\mathrm{eff}}^{2}}}\exp{(-z^{2}/2\sigma_{\mathrm{eff}}^{2})}\,,
\end{equation}
which describes the contribution of the telescope's beam and the projected emission of the infinitesimally thin disk.
The convolved emission profile has the form

\begin{eqnarray}
\label{convexp}
W_{\mathrm{exp}}(z)=\frac{w_{0}}{2}\exp{(-z^{2}/2\sigma_{\mathrm{eff}}^{2})}\cdot \nonumber\\
\left[\exp{\left(\frac{\sigma_{\mathrm{eff}}^{2}-zh}{\sqrt{2}\sigma_{\mathrm{eff}} h}\right)}^{2}\mathrm{erfc}\left(\frac{\sigma_{\mathrm{eff}}^{2}-zh}
{\sqrt{2}\sigma_{\mathrm{eff}} h}\right)\right) + \nonumber\\
\left(\exp{\left(\frac{\sigma_{\mathrm{eff}}^{2}+zh}{\sqrt{2}\sigma_{\mathrm{eff}} h}\right)}^{2}\mathrm{erfc}\left(\frac{\sigma_{\mathrm{eff}}^{2}+zh}
{\sqrt{2}\sigma_{\mathrm{eff}} h}\right) \right)]
\end{eqnarray}
in the case of an exponential intrinsic distribution \citep{sievers1988}, where $\mathrm{erfc}$ is the complementary error function, defined as
\begin{equation}
 \mathrm{erfc}\,(x)=\frac{2}{\sqrt{\pi}}\int_{x}^{\infty}\!\exp{(-r^{2})}\,\mathrm{d}r\,.
\end{equation}
For a Gaussian intrinsic distribution, the convolution yields \citep{dumke1994}
\begin{equation}
\label{convgauss}
 W_{\mathrm{Gauss}}(z)=\frac{w_{0}h}{\sqrt{2\sigma_{\mathrm{eff}}^{2}+h^{2}}}\exp{(-z^{2}/(2\sigma_{\mathrm{eff}}^{2}+h^{2}))}\,.
\end{equation}

The vertical radio scale heights $h$ can be determined by fitting one or two convolved exponential or Gaussian functions of the
form given in Eq.\,\ref{convexp} or Eq.\,\ref{convgauss} to the averaged radio intensities along each strip of a galaxy. The least-squares fits are made 
using the Levenberg-Marquardt method. Single boxes along a strip with mean values smaller than $2 \times$ the rms-noise
intensity level of the map are omitted by the fitting procedure. The quality of the fit along each strip is determined by a reduced $\chi^2$ test.

\section{Application to our sample and results}
\label{sec:results}
\subsection{Scale heights}
\label{subsec:scale heights}

The vertical scale heights were determined for all galaxies in our sample, as defined in Sect.~\ref{sec:sample} for C- and L-band. We applied the program
to the untapered but primary beam corrected maps of total intensity which had been cleaned with a robust 0 weighting, as described in Paper IV
\citep{wiegert+2015}. The strip width had to be adapted for each galaxy and band in such a way that first they are broader than the beam size
in order to gain sensitivity
and to have independent points between neighbouring strips, and second they are small enough to avoid strong intensity variation parallel
to the major axis which would increase the intensity dispersion within one box. The vertical box size (box height) was taken as $4 \arcsec$, which is
smaller than half the beam size. Otherwise, the sampling along the intensity profile would not be sufficient to fit the maximum of the vertical
intensity distribution well enough. This means that neighbouring points along one strip are not independent from each other. As the errors of
the individual points are only determined along the strip length, not the strip height (see Sect.~\ref{sec:method}), the selected box height does
not affect the errors of the individual points.

\begin{figure}
   \centering
   \includegraphics[width=0.95\columnwidth]{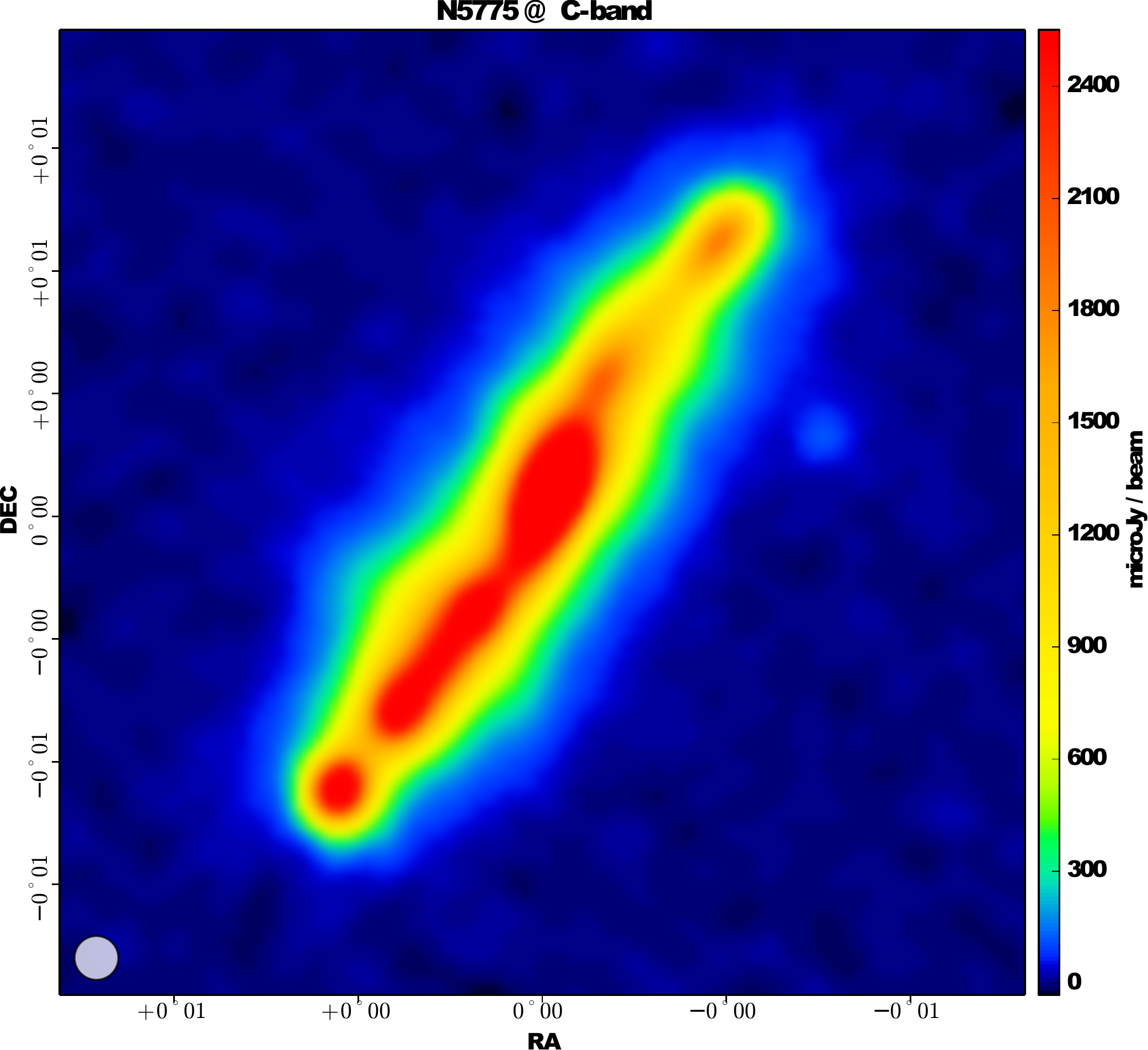}
      \caption{Total intensity colourized map of NGC~5775 at 6~GHz. The rms noise is $10\, \mu$Jy/beam with the beam size ($10.7 \arcsec$) given in bottom
      left corner.
              }
         \label{n5775map}
   \end{figure}

\begin{figure}
   \centering
   \includegraphics[width=0.95\columnwidth]{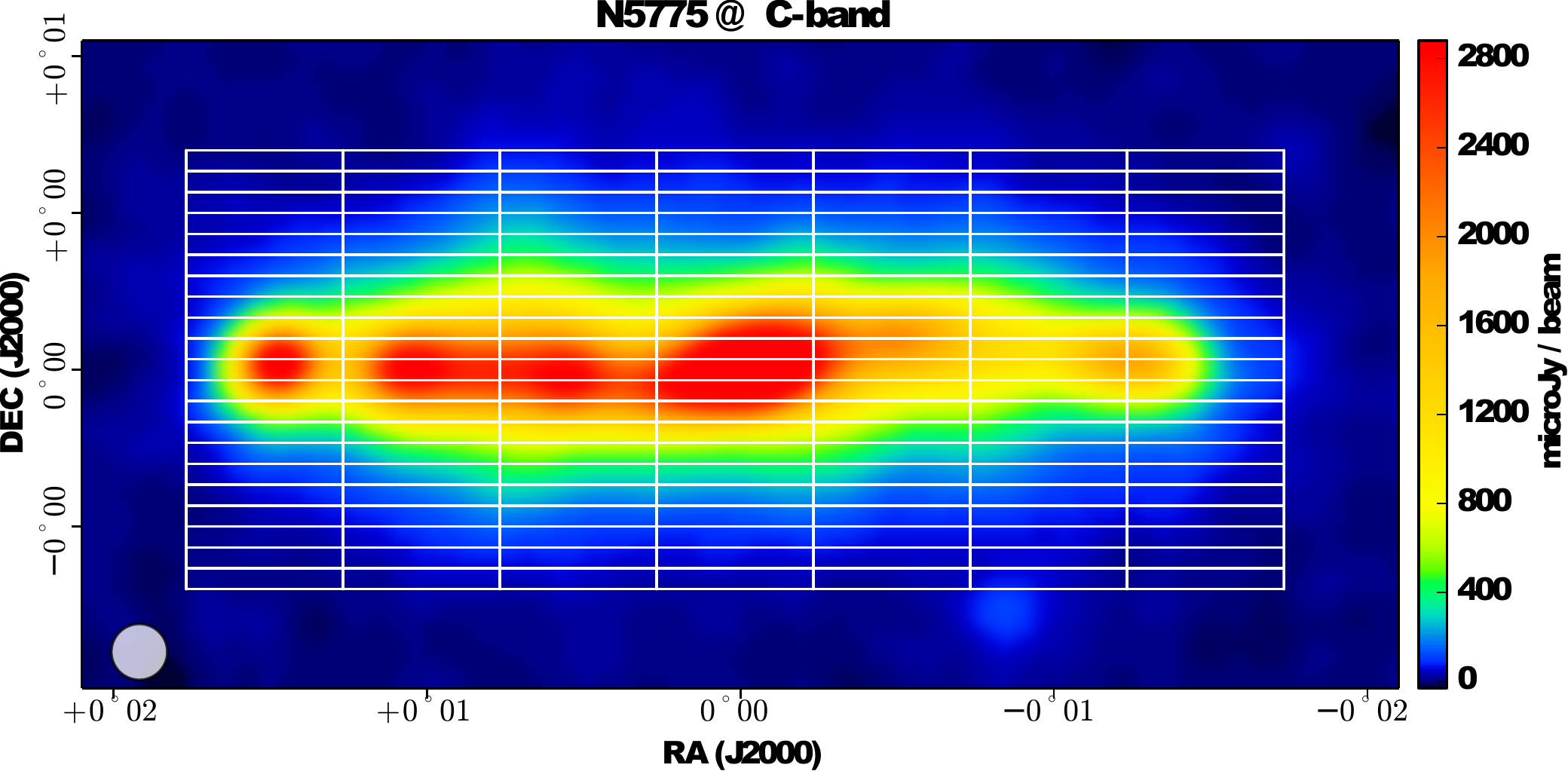}
      \caption{Rotated total intensity colourized map of NGC~5775 at 6~GHz with the strips and boxes for the scale height determination superimposed.
      In this case we used seven strips of $30\arcsec$ in width, with 21 boxes of $4\arcsec$ in height each.
      The beam size ($10.7 \arcsec$) is given in bottom left corner.
              }
         \label{n5775map.rotated}
   \end{figure}

\begin{figure*}
   \centering
   \includegraphics[width=1.95\columnwidth]{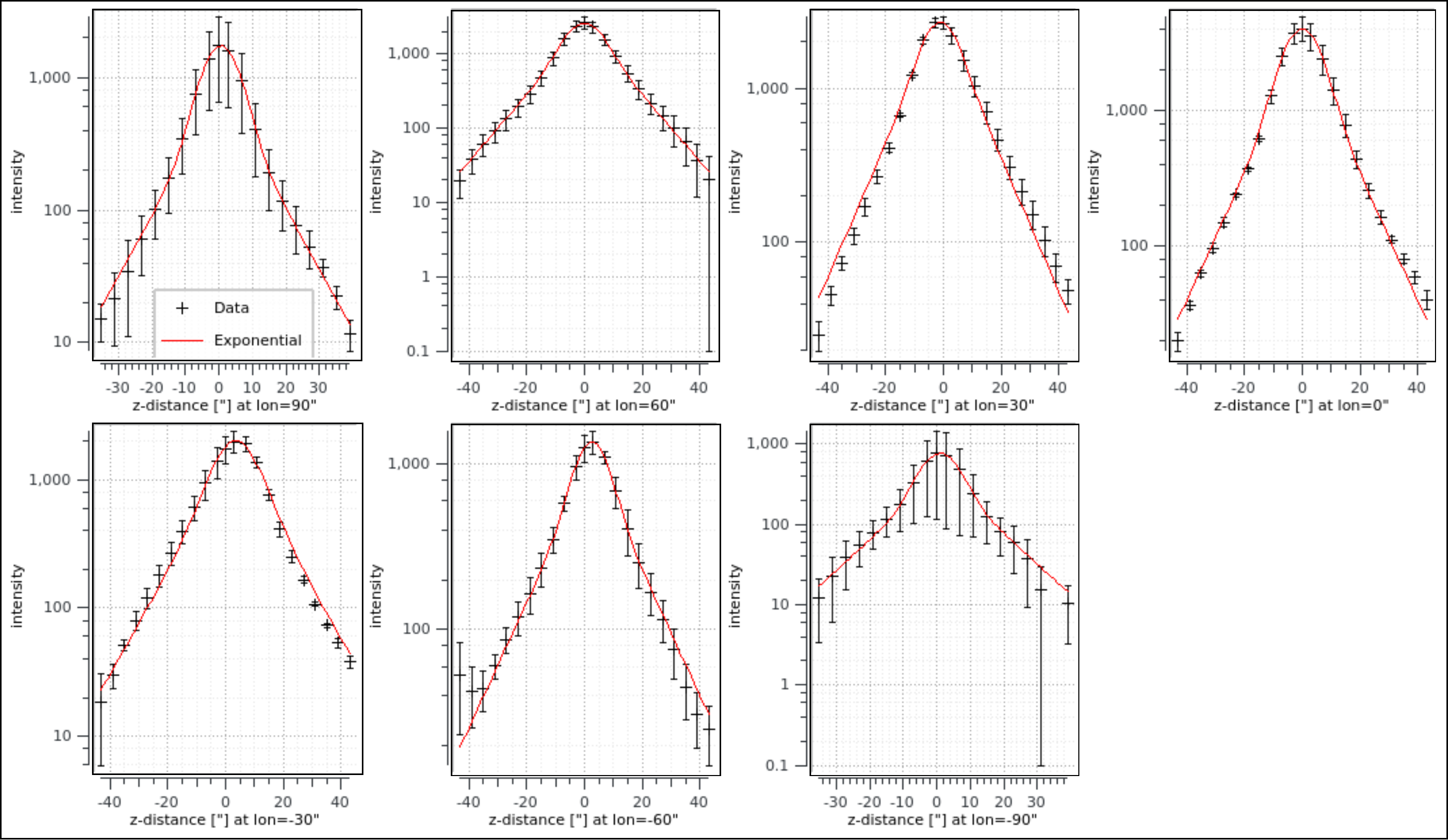}
      \caption{Averaged radio intensities of NGC~5775 within the boxes along the seven strips shown in Figure~\ref{n5775map.rotated} and their
      two-component exponential fits according to Eq.\,\ref{convexp}.
      }
         \label{n5775fits}
   \end{figure*}

Figure~\ref{n5775fits} presents the plots together with the results for each strip of our example galaxy NGC~5775 shown in Fig.~\ref{n5775map} and
Fig.~\ref{n5775map.rotated}. The parameters used for the fits of all galaxies in our sample are summarized in Table~\ref{fit parameter}.
Generally within our sample, the intensity distribution along the strips can be better fitted by exponential functions than by Gaussian functions.  First,
the data in each strip was fitted by a two component exponential function, as shown in Fig.~\ref{n5775fits}. Then the scale heights determined in each
strip of a galaxy were averaged together (except for some strongly deviating values with large errors which have been omitted). The averaged
scale heights for each galaxy together with their standard deviations at both frequency bands are listed in Table~\ref{fit results} as $h_0$ for the
thin disk exponential component and $h$ for the exponential thick disk / halo component. In the following, we will refer to $h$ as scale height of the halo.

\begin{table*}
      \caption[]{\label{fit parameter} Fit parameters}
      $$
         \begin{tabular}{lcccl}
            \hline
            \noalign{\smallskip}
              Source      & No. of  strips & strip width   &   No. of fit components   & remarks   \\
                        &  &  [\arcsec]  &  &     \\
            \hline
            \noalign{\smallskip}
            NGC~2820  &  5  &  30   &  2   &         \\
            NGC~3003  &  3  &  40   &  2   &         \\
            NGC~3044  &  3  &  40   &  2   &         \\
            NGC~3079  &  5  &  38   &  2   & AGN, central strip omitted \\
            NGC~3432  &  5  &  40   &  2   & tidal interaction        \\
            NGC~3735  &  5  &  26   &  2   &         \\
            NGC~3877  &  3  &  60   &  1   & same result as for 2 fit components \\
            NGC~4013  &  5  &  34   &  2   & AGN, central strip omitted for thin disk \\
            NGC~4157  &  5  &  40   &  2   &         \\
            NGC~4217  &  5  &  36   &  2   &         \\
            NGC~4302  &  5  &  44   &  2   & though AGN, no strip needed to be omitted       \\
            NGC~5775  &  7  &  30   &  2   &         \\
            U10288    &  4  &  37   &  1   & strip with background radio galaxy omitted,  \\
                      &     &       &      & same result as for 2 fit components \\
            \noalign{\smallskip}
            \hline
         \end{tabular}
      $$     

\end{table*}
\begin{table}
      \caption[]{\label{fit results} Fit results}
     $$
         \begin{tabular}{lccccc}
            \hline
            \noalign{\smallskip}
            \multicolumn{1}{c}{\textbf{Source}} &\multicolumn{2}{c}{\textbf{C-band D-array}} &\multicolumn{2}{c}{\textbf{L-band C-array}} \\
                        & $h_{0}\,^\mathrm{a}$ & $h$ & $h_{0}\,^\mathrm{a}$  & $h$  \\
                        & [\arcsec] & [\arcsec] & [\arcsec] & [\arcsec]  &\\
            \noalign{\smallskip}
            \hline
            \noalign{\smallskip}
            N2820  &  2.0 $\pm\,$ 1.6  &   9.1 $\pm\,$ 0.6   & 1.8 $\pm\,$ 0.5  & 11.6 $\pm\,$ 1.5  \\
            N3003  &  0.2 $\pm\,$ 3.1  &  10.4 $\pm\,$ 2.0   & 2.5 $\pm\,$ 2.5  & 29.7 $\pm\,$ 10.0 \\
            N3044  &  1.2 $\pm\,$ 1.8  &  10.5 $\pm\,$ 1.7   & 0.5 $\pm\,$ 0.7  & 13.6 $\pm\,$ 1.6  \\
            N3079  &  0.2 $\pm\,$ 0.1  &  15.4 $\pm\,$ 4.0   & 1.0 $\pm\,$ 1.4$\,^\mathrm{b}$ & 11.8 $\pm\,$ 1.9$\,^\mathrm{b}$ \\
            N3432  &  0.9 $\pm\,$ 0.9  &  15.1 $\pm\,$ 4.6   & 0.9 $\pm\,$ 0.9  & 20.3 $\pm\,$ 5.4  \\
            N3735  &  0.9 $\pm\,$ 1.2  &   7.0 $\pm\,$ 1.1   & 3.0 $\pm\,$ 0.4  &  8.1 $\pm\,$ 2.2  \\
            N3877  &         --        &  10.0 $\pm\,$ 1.8   &       --         & 12.5 $\pm\,$ 2.0  \\
            N4013  &  0.4 $\pm\,$ 0.3  &   8.2 $\pm\,$ 0.6   & 2.5 $\pm\,$ 1.1  & 13.2 $\pm\,$ 1.8  \\
            N4157  &  0.5 $\pm\,$ 0.6  &  13.0 $\pm\,$ 1.3   & 1.0 $\pm\,$ 1.4  & 14.4 $\pm\,$ 0.9  \\
            N4217  &  0.2 $\pm\,$ 0.1  &  10.8 $\pm\,$ 0.5   & 1.1 $\pm\,$ 0.7  & 10.6 $\pm\,$ 0.9  \\
            N4302  &  0.6 $\pm\,$ 0.6  &  11.9 $\pm\,$ 5.0   & 1.4 $\pm\,$ 1.3  & 10.7 $\pm\,$ 2.7  \\
            N5775  &  1.3 $\pm\,$ 1.3  &  10.4 $\pm\,$ 0.9   & 2.8 $\pm\,$ 1.2  & 14.2 $\pm\,$ 2.5  \\
            U10288  &            --     &   4.4 $\pm\,$ 0.1   &       --         &  5.1 $\pm\,$ 0.5  \\

            \noalign{\smallskip}
            \hline
         \end{tabular}
     $$
\begin{list}{}{}
\item[$^{\mathrm{a}}$] We note the limitations on the thin disk scale height as discussed in Sect.~\ref{subsec:scale heights}. They are not considered
further.
\item[$^{\mathrm{b}}$] The L-band values for NGC~3079 may be underestimated because of the poor quality of the map
(see Table~\ref{sample}).
\end{list}
   \end{table}

\begin{table*}
      \caption[]{\label{halo scale heights} Halo scale heights and normalized (halo) scale heights$\,^\mathrm{a}$}
     $$
         \begin{tabular}{lccccccc}
            \hline
            \noalign{\smallskip}
            \multicolumn{1}{c}{\textbf{Source}} &\multicolumn{1}{c}{\textbf{C-band D-array}} &\multicolumn{1}{c}{\textbf{L-band C-array}} &
            \multicolumn{1}{c}{\textbf{L-C-band}} &\multicolumn{2}{c}{\textbf{C-band}} &\multicolumn{2}{c}{\textbf{L-band}}\\
                         & $h$    & $h$  & scale height-  & \multicolumn{2}{c}{normalized scale height} & \multicolumn{2}{c}{normalized scale height}\\
                         & [kpc] & [kpc]  &  ratio &$\tilde{h}$ & $\Delta \tilde{h}$ & $\tilde{h}$ & $\Delta \tilde{h}$\\
            \noalign{\smallskip}
            \hline
            \noalign{\smallskip}
            NGC~2820  &  1.16 $\pm\,$ 0.07  &   1.48 $\pm\,$ 0.19  & 1.28 $\pm\,$ 0.18 & 5.32 & 0.42 & 6.14 & 0.85  \\
            NGC~3003  &  1.28 $\pm\,$ 0.25  &   3.66 $\pm\,$ 1.19  & 2.86 $\pm\,$ 1.08 & 6.84 & 1.38 & 19.06 & 6.28 \\
            NGC~3044  &  1.03 $\pm\,$ 0.11  &   1.33 $\pm\,$ 0.16  & 1.29 $\pm\,$ 0.21 & 5.45 & 0.64 & 6.72 & 0.88  \\
            NGC~3079  &  1.54 $\pm\,$ 0.04  &   1.18 $\pm\,$ 0.19$\,^\mathrm{b}$  & 0.77 $\pm\,$ 0.12$\,^\mathrm{a}$ & 6.09 & 0.35 & 5.67 & 0.95 \\
            NGC~3432  &  0.69 $\pm\,$ 0.21  &   0.93 $\pm\,$ 0.25  & 1.35 $\pm\,$ 0.55 & 5.61 & 1.73 & 6.64 & 1.82 \\
            NGC~3735  &  1.43 $\pm\,$ 0.23  &   1.66 $\pm\,$ 0.45  & 1.15 $\pm\,$ 0.37 & 5.13 & 0.86 & 4.78 & 1.32 \\
            NGC~3877  &  0.86 $\pm\,$ 0.16  &   1.08 $\pm\,$ 0.17  & 1.26 $\pm\,$ 0.31 & 4.67 & 0.90 & 5.57 & 0.92 \\
            NGC~4013  &  0.64 $\pm\,$ 0.05  &   1.03 $\pm\,$ 0.14  & 1.61 $\pm\,$ 0.25 & 3.81 & 0.35 & 6.02 & 0.88 \\
            NGC~4157  &  0.99 $\pm\,$ 0.10  &   1.09 $\pm\,$ 0.07  & 1.10 $\pm\,$ 0.13 & 5.13 & 0.58 & 5.59 & 0.46 \\
            NGC~4217  &  1.08 $\pm\,$ 0.05  &   1.06 $\pm\,$ 0.09  & 0.98 $\pm\,$ 0.09 & 4.78 & 0.32 & 4.34 & 0.43 \\
            NGC~4302  &  1.11 $\pm\,$ 0.47  &   1.01 $\pm\,$ 0.25  & 0.91 $\pm\,$ 0.45 & 5.31 & 2.26 & 4.43 & 1.12 \\
            NGC~5775  &  1.46 $\pm\,$ 0.13  &   1.98 $\pm\,$ 0.35  & 1.36 $\pm\,$ 0.27 & 4.55 & 0.46 & 5.50 & 1.01 \\
            UGC10288  &  0.73 $\pm\,$ 0.02  &   0.84 $\pm\,$ 0.08  & 1.15 $\pm\,$ 0.11 & 3.16 & 0.19 & 3.24 & 0.35 \\

            \noalign{\smallskip}
            \hline
         \end{tabular}
    $$
     
\begin{list}{}{}
\item[$^{\mathrm{a}}$] The normalized scale height is defined as $\tilde{h} = 100 \cdot h / d_{\rm r}$, where $h$ is the halo scale height and $d_{\rm r}$ the radio
    diameter at the same frequency.
\item[$^{\mathrm{b}}$] The L-band values for NGC~3079 may be underestimated because of the poor quality of the map
(see Table~\ref{sample}).
\end{list}
\end{table*}

In NGC~3079, the scale heights of the central strips have also been omitted in the averaging procedure because of the strong nuclear
source. Similarly, in UGC~10288 only strips which do not contain the background radio galaxy CHANG-ES~A \citep{irwin+2013} are considered.

Generally, the relative errors of the scale heights of the halo are  between 5\% and 20\%, whereas those of the thin
disk are much larger (50\% -- 100\%). Hence, the scale heights of the thin disk are not well determined. This is partly because
the expected scale heights for the thin disks are small ($2.5 \arcsec$ corresponds to about 300~pc for galaxies at a distance of 25~Mpc)
compared to the angular beam size of about $10 \arcsec$  of the observations (see Table~\ref{sample}). Another problem, especially
for the values for the thin disk, are the errors in the inclination, which are about $\pm\, 2\, \degr$. However, only the nearest galaxies with
distances $\lesssim 20$~Mpc have the smallest scale heights fitted for the thin disk in Table~\ref{fit results} with values corresponding to only
$50 \pm \, 50$~pc in C-band. This could hint at smaller scale heights for the thin disks than previously known.

The values for the halo scale heights
are in general $\gtrsim 10\, \arcsec$ (i.e. larger than the beam size), which is large enough to be reliably estimated. In this paper 
 from now on we only discuss the results for the halo. For two galaxies, NGC~3877 and UGC~10288, we found the same results for two fit
components. Hence, one component is enough to fit their vertical distributions. As these values are larger than expected for the thin
disk, we refer to them as halo scale heights.

The halo scale heights of the total radio emission are given in kpc in Table~\ref{halo scale heights}. The variation within our sample is  far greater
than that of the scale heights determined earlier for five large and more nearby edge-on galaxies  at 4.8~GHz ($1.8\, \pm \, 0.2$~kpc, \cite{krause2015}).
Our new values range between 0.7~kpc and 1.5~kpc in C-band, and between 0.9~kpc  and more than  3~kpc in L-band. The value for NGC~5775
in C-band (which is in both samples) is also smaller than the values given in \cite{soida+2011}. However, when corrected to the distance of 26.7~Mpc as taken in
\cite{soida+2011}, we derive $1.6 \pm 0.14$~kpc for the C-band halo scale height instead of $2.0 \pm 0.2$ as given in \cite{soida+2011}.
The remaining difference may be due to a higher angular resolution of our new maps ($11\arcsec$ instead of $16\arcsec$) and  to the more sophisticated
method of our new scale height determination which may be important for this small object (in angular size). We already note here
that other bright and more nearby galaxies (hence larger in angular and/or physical size) like NGC~891, NGC~4565, and NGC~4631 from the CHANG-ES
sample have larger halo scale heights when determined with the same new method (see Schmidt et al., in preparation, Mora-Partiarroyo et al., in preparation)
than the galaxies in our sample.

Our averaged values for the halo scale heights are $1.1\, \pm \, 0.3$~kpc in C-band and $1.4\, \pm \, 0.7$~kpc in L-band. As  the averaged L-band
value is also below the previous value determined at 4.8~GHz, our smaller values for the scale heights cannot be explained by missing large-scale flux density
as this affects  mainly the C-band values (see Sect.~\ref{subsec:missing spacings}).
\\

\subsection{Radial extents and possible effects of missing spacings}
\label{subsec:missing spacings}

Before we discuss our scale height values physically we have to make sure whether they suffer from missing large-scale flux density
in the map. As mentioned in Sect.~\ref{sec:sample}, we excluded galaxies which are larger in angular size than $6\arcmin$ from our sample.

\begin{figure}
   \centering
   \includegraphics[width=0.95\columnwidth]{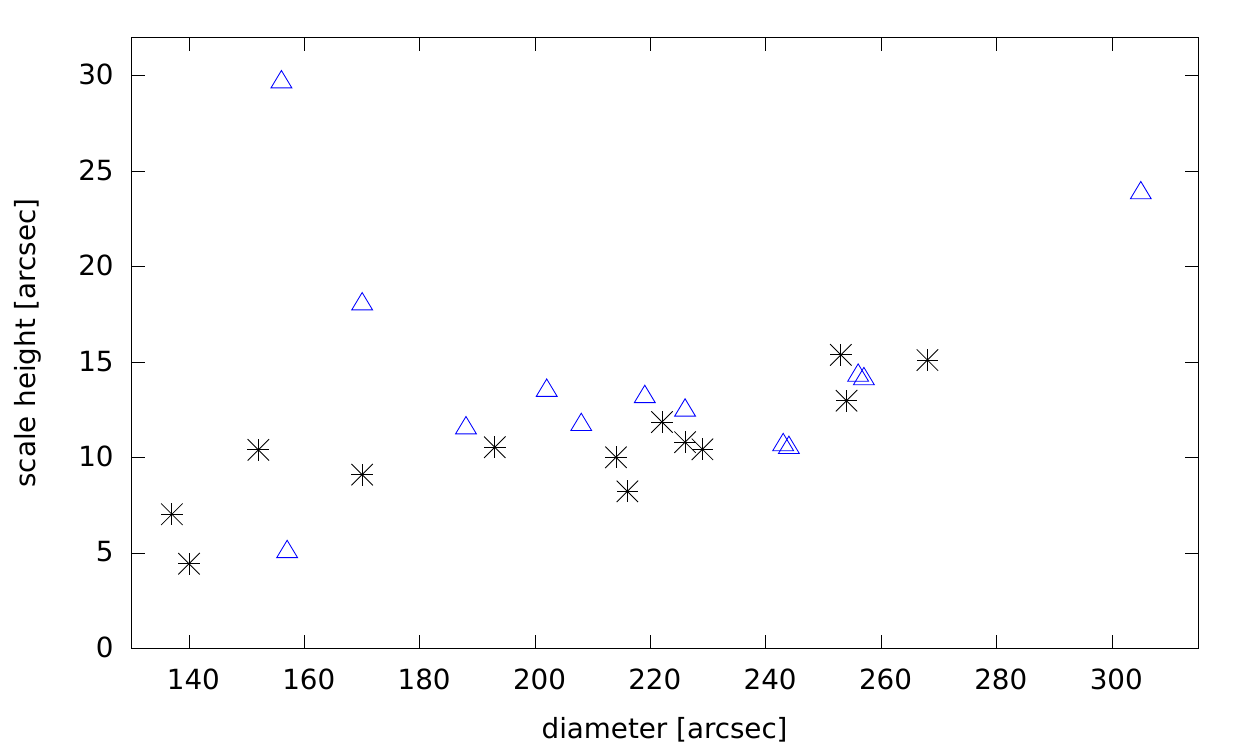}
      \caption{Halo scale heights vs. the galaxy's radio continuum diameter, both in arcsec in order to exclude errors in the
      distance determination. C-band data are represented by black asterisks, L-band data by blue triangles.
              }
         \label{arcsecdiameter}
   \end{figure}

Missing large-scale flux density for extended structures larger than 240\arcsec~is expected at C-band and only larger than 970\arcsec~at L-band.
For comparison, we determined the angular sizes of all sample galaxies in C- and L-band within the $2 \times \mathrm{rms}$ contours along the
major axis (see Cols. 2 and 3 (diameter) in Table~\ref{diameter}). While in L-band all sample galaxies are much smaller than 970\arcsec~in
angular size, in C-band there are several galaxies with angular diameters near or above 240\arcsec. Missing large-scale flux density
in these galaxies could lead to underestimated scale heights. In order to test this, we determined the radio diameter of each galaxy as the
angular size of the radio intensity within $2 \times$ the $\mathrm{rms}$  contours along the major axis (see Table~\ref{sample}) and plotted
the scale heights for C- and L-band against this quantity (both measured in arcsec), as shown in Figure~\ref{arcsecdiameter}.

\begin{table*}
      \caption[]{\label{diameter} Diameters}

     $$
         \begin{tabular}{lcccccc}
            \hline
            \noalign{\smallskip}
            Source & C-band & L-band & C-band & L-band & optical & L-C-band \\
                        & diameter & diameter & diameter & diameter & diameter$\,^\mathrm{a}$ & diameter \\
                        & [$\arcsec$] & [$\arcsec$] & [kpc] & [kpc] & [kpc] & ratio \\
            \noalign{\smallskip}
            \hline
            \noalign{\smallskip}
            NGC~2820  &  170 & 188 & 21.8  & 24.1  & 31.5 & 1.11 \\
            NGC~3003  &  152 & 156 & 18.7  & 19.2  & 44.3 & 1.02 \\
            NGC~3044  &  193 & 202 & 18.9  & 19.8  & 25.9 & 1.05 \\
            NGC~3079  &  253 & 208$\,^\mathrm{b}$ & 25.3  & 20.8$\,^\mathrm{b}$  & 46.2 & 0.82$\,^\mathrm{b}$ \\
            NGC~3432  &  268 & 305 & 12.3  & 14.0  & 13.5 & 1.14 \\
            NGC~3735  &  137 & 170 & 27.9  & 34.7  & 49.0 & 1.24 \\
            NGC~3877  &  214 & 226 & 18.4  & 19.4  & 26.3 & 1.05 \\
            NGC~4013  &  216 & 219 & 16.8  & 17.1  & 22.0 & 1.02 \\
            NGC~4157  &  254 & 256 & 19.3  & 19.5  & 31.9 & 1.01 \\
            NGC~4217  &  226 & 244 & 22.6  & 24.4  & 30.6 & 1.08 \\
            NGC~4302  &  222 & 243 & 20.9  & 22.8  & 26.5 & 1.09 \\
            NGC~5775  &  229 & 257 & 32.1  & 36.0  & 32.8 & 1.12 \\
            U10288    &  140 & 157 & 23.1  & 25.9  & 48.5 & 1.12 \\

            \noalign{\smallskip}
            \hline
         \end{tabular}
     $$
 \begin{list}{}{}
\item[$^{\mathrm{a}}$] Taken from \cite{wiegert+2015} based on \cite{tully1988}
\item[$^{\mathrm{b}}$] The L-band values for NGC~3079 may be underestimated because of the poor quality of the map
(see Table~\ref{sample}).
\end{list}
   \end{table*}

\begin{table*}
      \caption[]{\label{scalelength} Scale lengths and normalized scale lengths$\,^\mathrm{a}$}
     $$
         \begin{tabular}{lccccccccc}
            \hline
            \noalign{\smallskip}
            Source &  C-band & L-band &  L-C-band & \multicolumn{2}{c}{\textbf{C-band}} & \multicolumn{2}{c}{\textbf{L-band}} & C-band & L-band \\
                        & scale length$\,^\mathrm{b}$ & scale length & scale length- & \multicolumn{2}{c}{normalized scale length} & \multicolumn{2}{c}{normalized scale length}
                        & $h/l$ &  $h/l$ \\
                        & [kpc] & [kpc] & ratio & $\tilde{l}$ & $\Delta \tilde{l}$ & $\tilde{l}$ & $\Delta \tilde{l}$ & & \\
            \noalign{\smallskip}
            \hline
            \noalign{\smallskip}
            NGC~2820  & 3.1 $\pm\,$ 0.24 & 3.4 $\pm\,$ 0.22 &  1.09  & 14.22  & 1.31  & 14.11  & 1.15 & 0.37 $\pm\,$ 0.04 & 0.44 $\pm\,$ 0.06 \\
            NGC~3003  & 2.1 $\pm\,$ 0.17 & 2.5 $\pm\,$ 0.15 &  1.17  & 11.23  & 1.06  & 13.02  & 1.03 & 0.61 $\pm\,$ 0.13 & 1.46 $\pm\,$ 0.48 \\
            NGC~3044  & 1.7 $\pm\,$ 0.05 & 1.9 $\pm\,$ 0.04 &  1.14  &  8.99  & 0.50  &  9.60  & 0.53 & 0.61 $\pm\,$ 0.07 & 0.70 $\pm\,$ 0.09 \\
            NGC~3079  & 3.4 $\pm\,$ 0.25 & 3.1 $\pm\,$ 0.20 $\,^\mathrm{c}$ &  0.93$\,^\mathrm{c}$ &  13.44  & 1.21  & 14.90  & 1.20  & 0.45 $\pm\,$ 0.03 & 0.38 $\pm\,$ 0.07 \\
            NGC~3432  & 2.2 $\pm\,$ 0.13 & 2.8 $\pm\,$ 0.14 &  1.25  & 17.89  & 1.37  & 20.00  & 1.41 & 0.31 $\pm\,$ 0.10 & 0.33 $\pm\,$ 0.09 \\
            NGC~3735  & 4.7 $\pm\,$ 0.16 & 5.2 $\pm\,$ 0.18 &  1.11  & 16.85  & 1.02  & 14.99  & 0.90 & 0.30 $\pm\,$ 0.05 & 0.32 $\pm\,$ 0.09 \\
            NGC~3877  & 3.9 $\pm\,$ 0.57 & 5.5 $\pm\,$ 0.72 &  1.40  & 21.10  & 3.27  & 28.35  & 3.99 & 0.22 $\pm\,$ 0.05 & 0.20 $\pm\,$ 0.04 \\
            NGC~4013  & 2.7 $\pm\,$ 0.62 & 2.7 $\pm\,$ 0.62 &  1.00  & 16.07  & 3.77  & 15.79  & 3.72 & 0.24 $\pm\,$ 0.06 & 0.38 $\pm\,$ 0.10 \\
            NGC~4157  & 5.9 $\pm\,$ 0.46 & 6.5 $\pm\,$ 0.53 &  1.09  & 30.57  & 2.86  & 33.33  & 3.21 & 0.17 $\pm\,$ 0.02 & 0.17 $\pm\,$ 0.22 \\
            NGC~4217  & 4.9 $\pm\,$ 0.33 & 5.0 $\pm\,$ 0.26 &  1.04  & 21.68  & 1.80  & 20.49  & 1.47 & 0.22 $\pm\,$ 0.02 & 0.21 $\pm\,$ 0.02 \\
            NGC~4302  & 4.0 $\pm\,$ 0.34 & 4.4 $\pm\,$ 0.30 &  1.12  & 19.14  & 1.87  & 19.30  & 1.61 & 0.28 $\pm\,$ 0.12 & 0.23 $\pm\,$ 0.06 \\
            NGC~5775  & 11.1 $\pm\,$ 1.54 & 11.3 $\pm\,$ 1.16 & 1.02 & 34.58  & 5.10  & 31.39  & 3.58 & 0.13 $\pm\,$ 0.02 & 0.18 $\pm\,$ 0.04 \\

            \noalign{\smallskip}
            \hline
         \end{tabular}
     $$
 \begin{list}{}{}
\item[$^{\mathrm{a}}$] The normalized scale length is defined as $\tilde{l} = 100 \cdot l / d_{\rm r}$, where $l$ is the scale length and $d_{\rm r}$ is the radio diameter
    at the same frequency.
\item[$^{\mathrm{b}}$] No reasonable fit for the scale length could be made for UGC~10288 because of the bright radio galaxy CHANG-ES A located asymmetrically behind the disk.
\item[$^{\mathrm{c}}$] The L-band values for NGC~3079 may be underestimated because of the poor quality of the map (see Table~\ref{sample}).
\end{list}
   \end{table*}

   In contrast to what would be expected for missing large-scale structure (namely a decrease in scale heights with diameter when both are
   measured in arcsec), the scale heights determined at C-band increase with diameter. At
   larger diameters  the scale heights at L-band also increase, similarly to the C-band data. Hence, we conclude that we have no obvious missing
   spacing problem for our sample galaxies. However, we cannot exclude them. In this case the C-band scale heights would give lower limits for
   the radio scale heights.

\section{Radial scale lengths, distances, and diameters}
\label{sec:scale length}

In addition to the vertical scale heights we also determined the radial scale lengths of all sample galaxies. As learned from the analysis of stellar
disks in edge-on galaxies, the line-of-sight integration of an exponential distribution in radius results in a Bessel function that can be closely approximated
 by an exponential with the same scale length \citep{kruit+1981}. The evaluation was done with the same task in the NOD3
package \citep{mueller+2017}, similar to that of the scale height determination (see Sect.~\ref{sec:method}) except that the strip is now
parallel to the major axis with a width of about $ 1-2.5 \times $HPBW. The distribution is again corrected for the beam size. However, no correction for the
inclination is necessary. We fitted a one-component exponential fit to the radial intensity distribution. For the galaxies with a strong nuclear source,
we made a two-component fit (for NGC~3079) or subtracted the central source before the fitting (for NGC~4013). No reasonable fit for the scale length
could be made for UGC~10288 because of the bright radio galaxy CHANG-ES A located asymmetrically behind the disk. The scale lengths for C- and L-band are given in
Table~\ref{scalelength} together with their L- to C-band ratios. They vary over a much larger range than the scale heights.

\begin{figure}
   \centering
   \includegraphics[width=0.95\columnwidth]{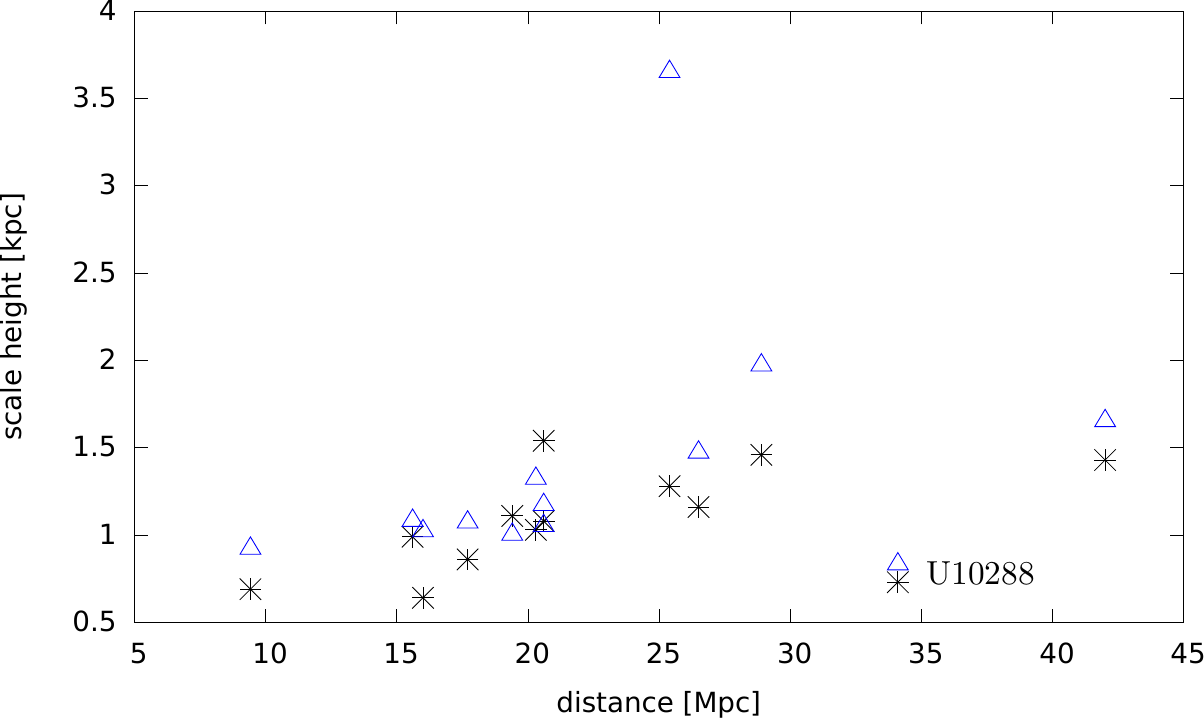}
      \caption{Halo scale heights vs. the assumed distances. C-band data are represented by black asterisks,
      L-band data by blue triangles.
              }
         \label{distance-height}
   \end{figure}

\begin{figure}
   \centering
   \includegraphics[width=0.95\columnwidth]{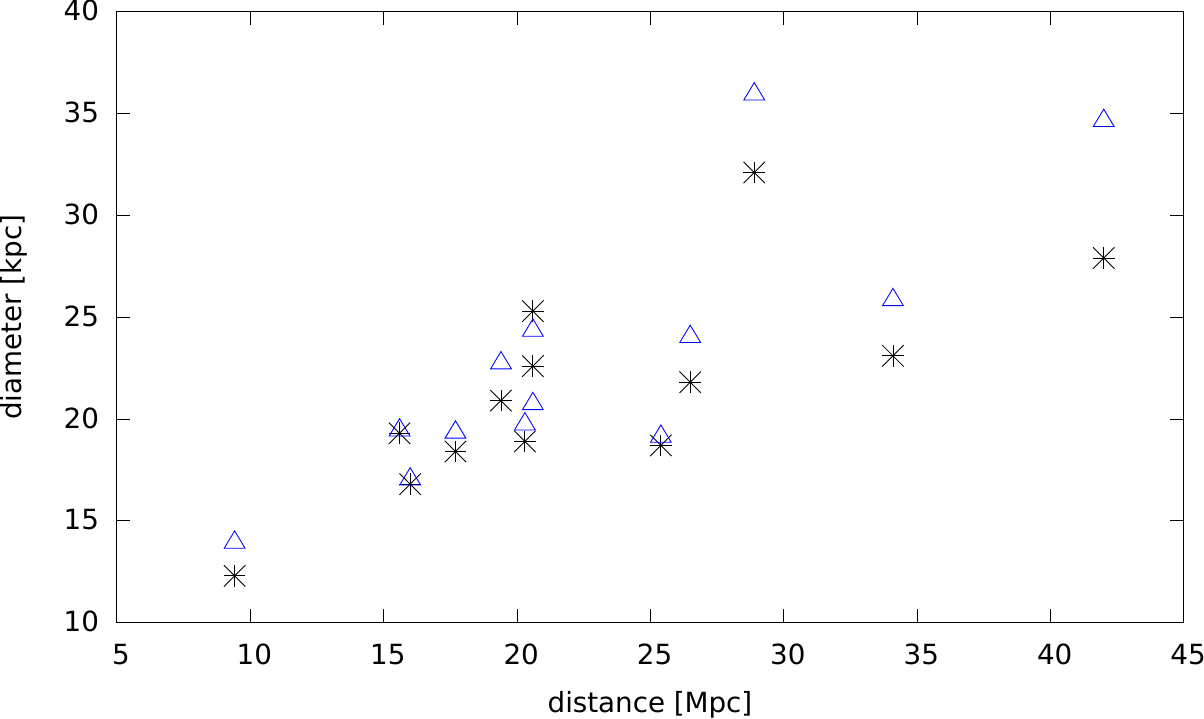}
      \caption{Observed radio diameter vs. the assumed distance. C-band data are represented by black asterisks,
      L-band data by blue triangles.
              }
         \label{distance-dia}
   \end{figure}

\begin{figure}
   \centering
   \includegraphics[width=0.95\columnwidth]{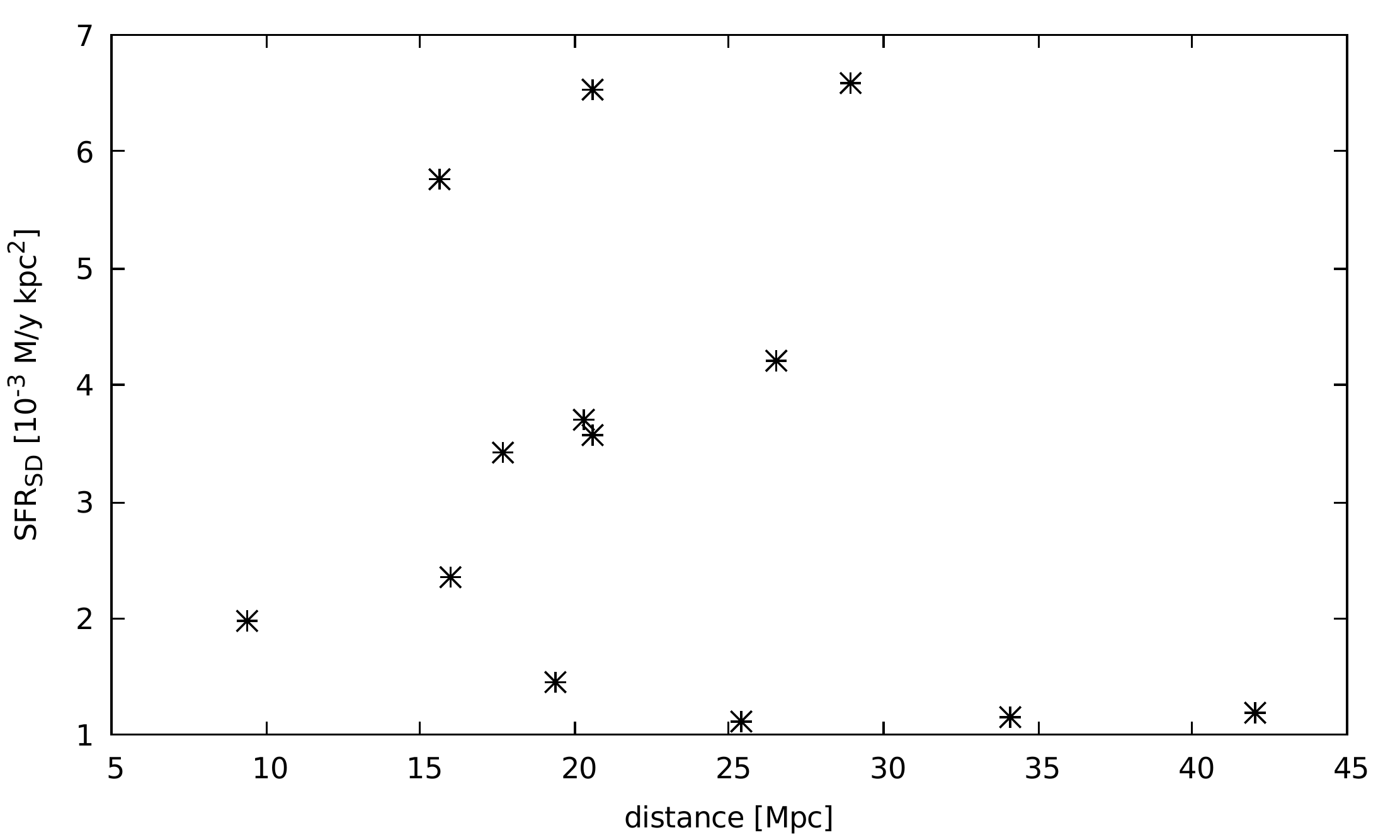}
      \caption{Star formation rate surface density SFR$_\mathrm{SD}$ vs. the distances of the sample galaxies.
              }
         \label{distance-SFRSD}
   \end{figure}

The distances used (as given in Table~\ref{sample}) are taken from CHANGES Paper IV \citep{wiegert+2015}. In order to test all distances we plotted
the scale heights as well as the diameters (see Sect.~\ref{subsec:missing spacings}) with distance in Figure~\ref{distance-height} and
Figure~\ref{distance-dia}. The indication in Figure \ref{distance-dia} that the radio diameters $d_r$ increase with distance is understandable as a result of
the interplay of the selection criteria of the CHANG-ES galaxies. The sample is flux density limited (at the lower end) and fixed in angular size
range (see Sect.~\ref{sec:sample}). This range is even smaller for our
subsample ($4\arcmin < d < 6\arcmin$) and means that our sample is biased such that large but nearby galaxies are omitted (due to the upper
size limit), while with increasing distances only larger galaxies are extended enough to fulfil the lower size limit. On the other hand, their flux
densities decrease quadratically with distance, which restricts the detectability of the outer parts of their disks to a given sensitivity of
the observations. The lower limit for the flux density, however, prefers brighter galaxies. Even so, we stress that our sample
also contains galaxies with low star formation rate surface density (SFR$_\mathrm{SD}$) (taken from \cite{wiegert+2015}) at large distances, as shown
in Figure~\ref{distance-SFRSD}.

The halo scale heights (measured in kpc) increase slightly with distance as shown in Figure~\ref{distance-height}. The only exception is
UGC~10288, which is not surprising since the CHANG-ES observations revealed the presence of a strong,
double-lobed extragalactic radio source (CHANG-ES A) located almost directly behind this edge-on galaxy \citep{irwin+2013}. UGC~10288 would have fallen well
below the CHANG-ES flux density cut-off had it been considered without the dominant contribution of what has now been shown to be the background source CHANG-ES A.
The values for UGC~10288 presented in this paper are determined after the separation of the two objects.

\section{Magnetic field strengths}
\label{sec:field strength}

We determined the galaxy-averaged total magnetic field strength in all galaxies from the C-band images with the assumption of energy equipartition between the magnetic
field and the cosmic rays according to \cite{beck+2005}. The line of sight (L) through the galaxy was estimated using 80\%
of the measured diameter of each galaxy within $5 \times$ the rms contour of the total power image. The non-thermal spectral index $\alpha_n$
was estimated from the flux densities given in Paper IV \citep{wiegert+2015} (except for NGC~3079) with the assumption that the thermal
fraction at C-band is 20\% and negligible at L-band. The determined values for $\alpha_n$ are all close to 1 (see Table~\ref{bfeld}). As the
determined spectral index value for NGC~3079 is strongly influenced by the nucleus and its outflow, we simply assumed a value of 1.0 for this galaxy.
We further assumed a proton-to-electron number density $\rm {K}=100$.

We determined the average magnetic field strengths for the total galaxy B$_\mathrm{t}$ and for the disk B$_\mathrm{t(disk)}$ separately.
For the calculation of B$_\mathrm{t}$ we used the intensity at C-band averaged over the entire galaxy within $5 \times$ the rms~contour,
whereas for B$_\mathrm{t(disk)}$ we averaged the radio intensity over that same extent projected along the minor axis at the inclination
of the galaxy. For our sample, the uncertainty in the non-thermal spectral index $\alpha$ is the dominant contribution to the error in the magnetic field
strengths which is $\lesssim 1 \, \mu$G for $\Delta \alpha \lesssim 0.1$. The adopted values of L, non-thermal spectral indices, and determined magnetic
fields strengths are summarized in Table~\ref{bfeld}
together with the star formation rates (SFR) and the star formation rate surface density (SFR$_\mathrm{SD}$) taken from Paper IV \citep{wiegert+2015}.

The averaged total field strengths B$_\mathrm{t}$ for all galaxies are in the narrow range of $9-11~\mu$G, except for the very faint
galaxy UGC~10288. This can be easily understood by our selection criteria of the sample galaxies with respect to the size ($d_{25} > 4\arcmin$ and $d_r < 5\arcmin$)
and flux density ($ S_{1.4}> 23$~mJy) (see Sect.~\ref{sec:sample}) and the fact that the equipartition magnetic field strengths depends on the radio surface
brightness of the galaxies with a small exponent of about 1/4. Hence, our galaxy sample just includes the spiral galaxies with the highest surface brightness, except for
UGC~10288.
The averaged magnetic field strengths of the disk B$_\mathrm{t(disk)}$ are slightly larger and vary over a somewhat wider range, between $(10 - 15)~\mu$G,
again except for UGC~10288. In contrast, the SFR of the sample galaxies is spread over quite a large range ($(0.5 - 5.3)\, \mathrm{M}_{\odot}\,$yr$^{-1}$), similar to
SFR$_\mathrm{SD}$ which is in the range of $(1.1 - 6.6) \, 10^{-3}\, \mathrm{M}_{\odot}\,$yr$^{-1}$~kpc$^{-2}$ \citep{wiegert+2015}.

\begin{table*}
      \caption[]{\label{bfeld} Magnetic field strengths, star formation rates$\,^\mathrm{a}$, and total mass surface densities}
     $$
         \begin{tabular}{lccccccc}
            \hline
            \noalign{\smallskip}
                        Source &  L  & $\alpha_{\rm nt}$  & B$_\mathrm{t}$ & B$_\mathrm{t(disk)}$ & SFR & $\rm{SFR}_\mathrm{SD}$ & MSD \\
                        & [kpc]  &  & [$\mu$G]  & [$\mu$G] & [M$_{\odot}\,$yr$^{-1}$]  & [$10^{-3}$M$_{\odot}\,$yr$^{-1}$ kpc$^{-2}$] & [$ 10^7 \, $M$_{\odot} \, $kpc$^{-2}$ ]
                        \\
            \noalign{\smallskip}
            \hline
            \noalign{\smallskip}
            N2820  &  16 & 1.04 & 10.1 & 12.6 & 0.62 & 4.20 & 9.51 \\
            N3003  &  12 & 1.05 & 10.0 & 11.8 & 0.67 & 1.11 & 3.65 \\
            N3044  &  14 & 0.93 & 10.7 & 13.8 & 0.95 & 3.70 & 12.26 \\
            N3079  &  18 & $1.0^\mathrm{b}$  & 11.4 &  $--^\mathrm{c}$  & 3.46 & 6.54 & 7.01 \\
            N3432  &  9  & 1.03 & 10.6 & 12.3 & 0.15 & 1.97 & 10.84 \\
            N3735  &  21 & 1.05 & 10.7 & 13.4 & 1.10 & 1.20 & 12.72 \\
            N3877  &  13 & 1.06 &  9.7 & 10.5 & 0.92 & 3.43 & 10.13 \\
            N4013  &  13 & 0.99 & 10.0 & 12.2 & 0.48 & 2.35 & 16.24 \\
            N4157  &  14 & 1.07 & 11.3 & 13.7 & 1.25 & 5.77 & 12.52 \\
            N4217  &  17 & 1.03 & 10.7 & 13.3 & 1.53 & 3.57 & 14.59 \\
            N4302  &  15 & 1.00 &  9.7 & 11.3 & 0.53 & 1.45 & 14.38 \\
            N5775  &  24 & 1.09 & 11.5 & 14.8 & 5.28 & 6.58 & 14.56 \\
            U10288 &  16 & 0.96 &  6.3 &  6.7 & 0.41 & 1.15 & 6.36 \\

            \noalign{\smallskip}
            \hline
         \end{tabular}
     $$
\begin{list}{}{}
\item[$^{\mathrm{a}}$] The star formation rates are taken from Paper IV \citep{wiegert+2015} and are corrected for AGN contamination.
\item[$^{\mathrm{b}}$] As the determined spectral index of NGC~3079 is strongly influenced by the nucleus and its outflow, we adopted this value.
\item[$^{\mathrm{c}}$] This value could not be determined because of the strong background source.
\end{list}
   \end{table*}

\section{Frequency dependence of radio scale heights}
\label{sec:frequency}
\begin{figure}
   \centering
   \includegraphics[width=0.85\columnwidth]{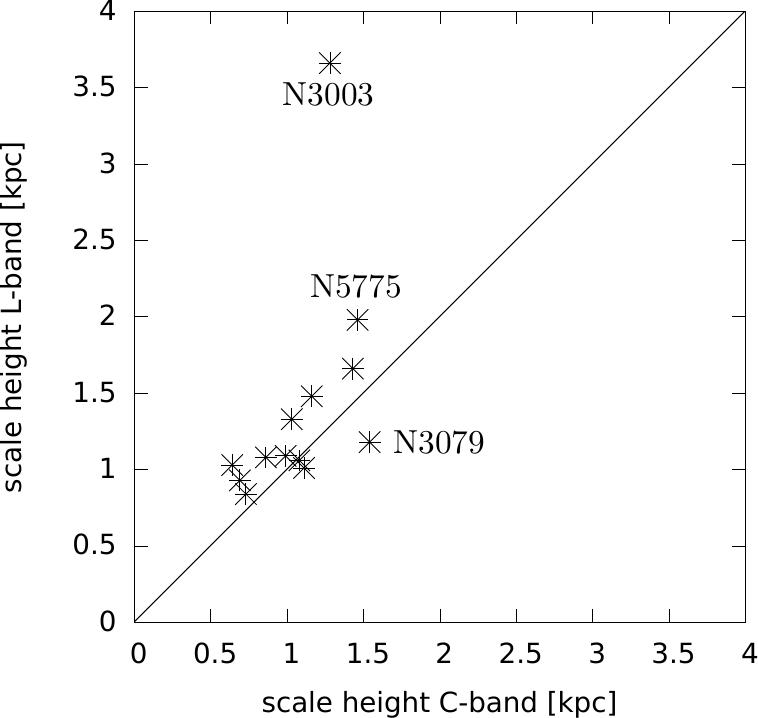}
      \caption{Radio scale heights in C- and L-band.
              }
         \label{L-C-scaleheights}
   \end{figure}

The shape of a radio halo and the variation of the radio scale heights with frequency can tell us a lot about the possible loss or
transport mechanisms of the CRE in the halo (see Sect.~\ref{sec:introduction}). Hence, we plotted the C- and L-band scale heights
in Fig.~\ref{L-C-scaleheights}. Most of the sample galaxies are located near to the one-to-one-line with one strong outlier, NGC~3003. The L-band
values for NGC~3079 may be underestimated because of the poor quality of the map (see Table~\ref{sample}). Hence, altogether the scale
heights at L-band are larger or about equal to those at C-band.

The ratio of the radio scale heights as measured in L- and C-band are given in Table~\ref{halo scale heights} together with its errors as calculated by error
propagation. The uncertainty in NGC~3079 due to the high noise error of the L-band map is not included. The ratio of the scale lengths at both bands
is about 1 within 10\% for most of the galaxies (see Table~\ref{scalelength}).

NGC~3003 reveals the largest difference between the two bands. However, its
inclination is not well determined and was very difficult to estimate even with our scale height fits (as described in Sect.~\ref{sec:sample}). It is
described as a disturbed galaxy with asymmetric spiral arms \citep{karthick+2014} with indications for a starburst superwind \citep{hoopes+1999}. Hence,
the inclination may be smaller than $84\degr$ and we cannot fully trust our scale height values for this galaxy.

For both synchrotron loss and escape-dominated halos the CR transport from the disk into the halo is assumed to be due to diffusive or convective propagation which
lead to different frequency dependencies of the radio scale heights. In a synchrotron energy loss-dominated halo the outer halo boundary is observationally
determined where the CRE have lost their energy by a sufficient amount that its radiation is below the detection limit of our observations. In an
escape-dominated halo the halo boundary is observed where the number of CRE and the magnetic field strengths are too small to emit detectable
radio emission, which is also
dependent on the sensitivity of the observations. In practice both processes can occur simultaneously along the line of sight. In the following, we
estimate the frequency and magnetic field strength dependencies of the synchrotron energy loss-dominated halo (a) and the escape-dominated halo (b), each for
diffusive and convective CRE propagation.

(a) Following \cite{heesen+2009}, based on e.g. \cite{longair2011}, the synchrotron cooling timescale for CRE is given by

\begin{equation}
\label{tsyn}
 t_{\rm syn}(\mathrm{yr})\, = \,8.35 \cdot 10^9 \, E(\mathrm{GeV})^{-1}B_{\bot}(\mathrm{\mu G})^{-2}
,\end{equation}
where E is the CRE energy and B$_{\bot}$ the field strength perpendicular to the CRE motion where we can assume $ \mathrm{B_\bot} = \sqrt{2/3} \, \mathrm {B_t}$ for an
isotropic distribution of magnetic field orientations. Synchrotron emission at frequency $\nu$ comes from CRE with the average energy

\begin{equation}
\label{energy}
 E(\mathrm{GeV}) \, = \, (\nu / 16.1 \, \mathrm{MHz})^{1/2}\,B_{\bot}(\mathrm{\mu G})^{-1/2}.
\end{equation}
For a field strength of $10~\mu$G this is about $6$~GeV at C-band and $3$~GeV at L-band.

Using Eq.\,\ref{tsyn} and Eq.\,\ref{energy} the synchrotron timescale for the CRE can be expressed by
$t_{\rm CRE} = t_{\rm syn} \, \propto \, B_{\bot}^{-3/2} \nu^{-1/2}$. The CRE diffusion length $l_{\rm diff} \, \propto \, (D \, t_{\rm CRE})^{1/2}$, where $D$ is the
diffusion coefficient. \cite{mulcahy+2016} found no evidence for an energy dependence of $D$ in M~51 in a similar CRE energy range of our observations, hence we obtain
for a synchrotron energy loss-dominated halo

\begin{equation}
\label{diffusion.notenergydep}
l_{\rm diff} \, \propto \, B^{-3/4} \,\nu^{-1/4},
\end{equation}
while \cite{murphy2009} suggested $D \, \propto \, E^{1/2}$, leading to
$D \, \propto \, B^{-1/4} \, \nu^{1/4}$ \citep{taba+2013} and hence to

\begin{equation}
\label{diffusion.energydep}
l_{\rm diff} \, \propto \, B^{-7/8} \, \nu^{-1/8}.
\end{equation}

Synchrotron intensity depends on the product of CRE number density and the magnetic field strength to the power of about 2. Consequently, the exponential scale
height of the synchrotron emission $h_{\rm z}$ depends on the exponential CRE scale height $h_{\rm CRE}$ and the exponential scale height of the magnetic field $h_{\rm B}$ as

\begin{equation}
h_{\rm z} \, \simeq \, h_{\rm CRE} / ((2\,h_{\rm CRE}/h_{\rm B}) \, +1))
\label{eq:scaleheight}
,\end{equation}
where $h_{\rm CRE}$ is given by the diffusion length $l_{\rm diff}$.
Magnetic fields are expected to extend far above the disk, much further than CRE can
propagate by diffusion, so that $l_{\rm diff} \ll h_{\rm B}$ and $h_{\rm z} \approx l_{\rm diff}$.

In the above case of diffusion in a synchrotron energy loss-dominated halo we get

\begin{equation}
 h_{\rm z \mathrm {(L-band)}} \, \approx \, 1.2 \, h_{\rm z \mathrm {(C-band)}}
\end{equation}
if $D$ is energy dependent ($D \, \propto \, E^{1/2}$), and

\begin{equation}
 h_{\rm z \mathrm {(L-band)}} \, \approx \, 1.4 \, h_{\rm z \mathrm {(C-band)}}
\end{equation}
if $D$ is not energy dependent.

For convection the CRE propagation length $l_{\rm con}$ is determined by the convection velocity, assumed to be the wind velocity $\mathrm{v_{wind}}$,
hence

\begin{equation}
\label{convection}
l_{\rm con} \, = \, \mathrm{v_{wind}} \, t_{\rm syn} \, \propto \, B^{-3/2} \, \nu^{-1/2}.
\end{equation}

In the case of streaming, $\mathrm{v_{wind}}$ has to be replaced by the Alfv\'{e}n velocity $\mathrm{v_A}$. Both velocities are expected to increase with distance from the
galactic plane, but the detailed dependencies could not yet be measured. Assuming that $l_{\rm con}$ is proportional to the radio scale height h$_{z}$, we obtain for
convection in a synchrotron energy loss-dominated halo

\begin{equation}
 h_{\rm z \mathrm {(L-band)}} \, \approx \, 2.0 \, h_{\rm z \mathrm {(C-band)}}.
\end{equation}

(b) In the case of an escape-dominated halo (i.e. $t_{\rm esc} \ll t_{\rm syn}$) the CRE confinement time is expected to be equal to the CRE escape time $t_{\rm esc}$
and to be roughly constant.
Here, the CRE scale height $h_{\rm CRE}$ is probably determined by energy equipartition between cosmic rays and magnetic fields, so that $h_{\rm CRE} = h_{\rm B}/2$ 
and, according to Eq.\,\ref{eq:scaleheight}, $h_{\rm z} = h_{\rm CRE}/2 = h_{\rm B}/4$. Hence, $h_{\rm CRE}$ is larger than in 
the case of a synchrotron energy loss-dominated halo.

Using the different energy dependencies of the diffusion length $D$ as given in Eq.\,\ref{diffusion.energydep} and
Eq.\,\ref{diffusion.notenergydep}, the exponential synchrotron scale height only weakly depends on the observed frequency and for an escape-dominated
halo with diffusion we obtain

\begin{equation}
 h_{\rm z \mathrm {(L-band)}} \, \approx \, (1.1 - 1.2) \, h_{\rm z \mathrm {(C-band)}}.
\end{equation}

For convection, the wind velocity may only depend on the magnetic field and not on the observed frequency, hence we conclude for an escape-dominated
halo with convective propagation

\begin{equation}
 h_{\rm z \mathrm {(L-band)}} \, \approx \, h_{\rm z \mathrm {(C-band)}}.
\end{equation}

\begin{figure}
   \centering
   \includegraphics[width=0.9\columnwidth]{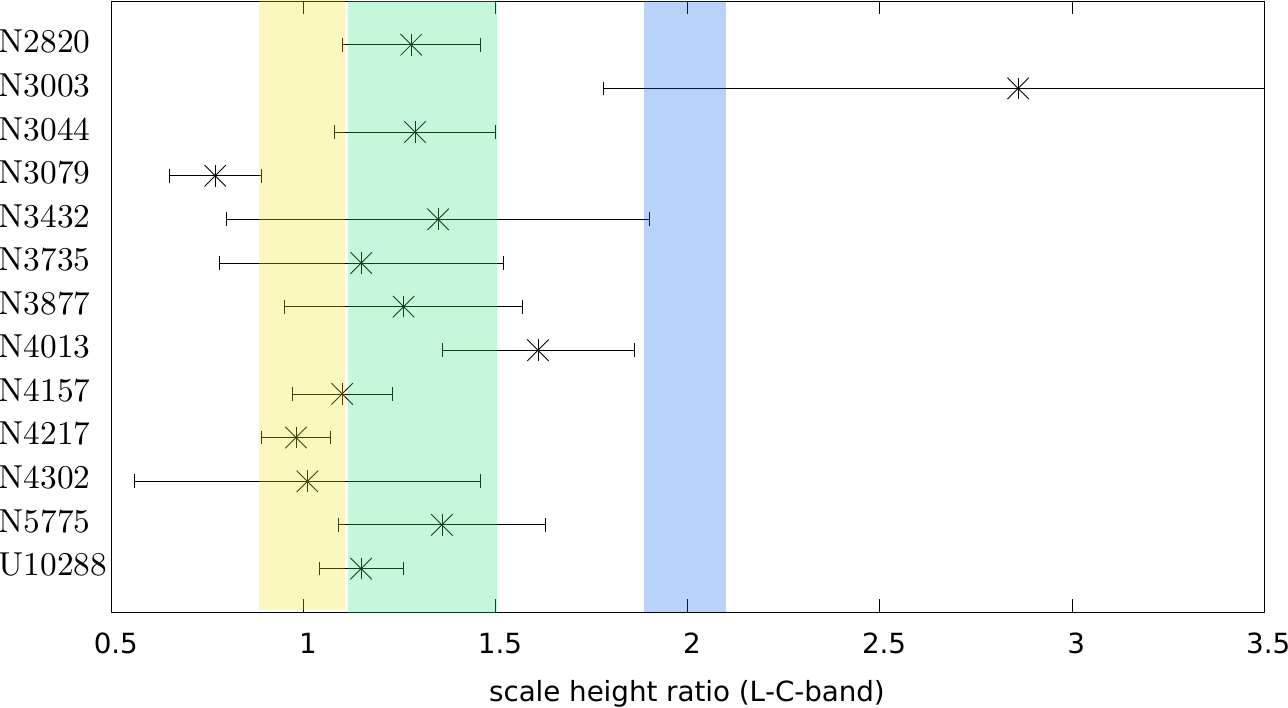}
      \caption{Ratio of the scale heights in L- and C-band with errors for all galaxies. The coloured areas indicate the expected ranges for an
      escape-dominated halo with convection (yellow), a synchrotron energy loss-dominated or escape-dominated halo with diffusion (green), and a synchrotron
      loss-dominated halo with convection (blue).
              }
         \label{ratio}
   \end{figure}

From the estimates above we see that a distinction between diffusion in a synchrotron energy loss-dominated halo and diffusion in an escape-dominated halo using
the frequency dependence of the radio scale heights alone is hardly possible. The ratio of the determined exponential scale heights for the halo in L-band and C-band 
are given in Table~\ref{halo scale heights} and graphically presented in Fig.~\ref{ratio}, together with their errors.

Remarkably, the ratios for all galaxies  range between 1 and 2, except for the outliers NGC~3003 and NGC~3079,  mentioned already
at the beginning of this section. The coloured areas in Fig.~\ref{ratio} indicate the expected ranges for an escape-dominated halo with convection (in
yellow), a synchrotron energy loss-dominated or escape-dominated halo both with diffusion (in green), and a synchrotron energy loss-dominated halo with convection (in blue).
Including the errors, only NGC~3003 could be a galaxy with a synchrotron energy loss-dominated halo with convective propagation. The others seem to have either
escape-dominated halos or synchrotron energy loss-dominated halos with diffusion, between which a separation on the basis of the ratios alone is difficult
due to their errors. Nevertheless, from the frequency analysis the clearest candidates for having an escape-dominated halo with convective propagation are NGC~4157, NGC~4217,
and NGC~4302; the others are in the range of a synchrotron energy loss-dominated or escape-dominated halo, both with diffusion, except NGC~3003 and NGC~3079.

\section{Correlations}
\label{sec:correlations}

\begin{figure}
   \centering
   \includegraphics[width=0.95\columnwidth]{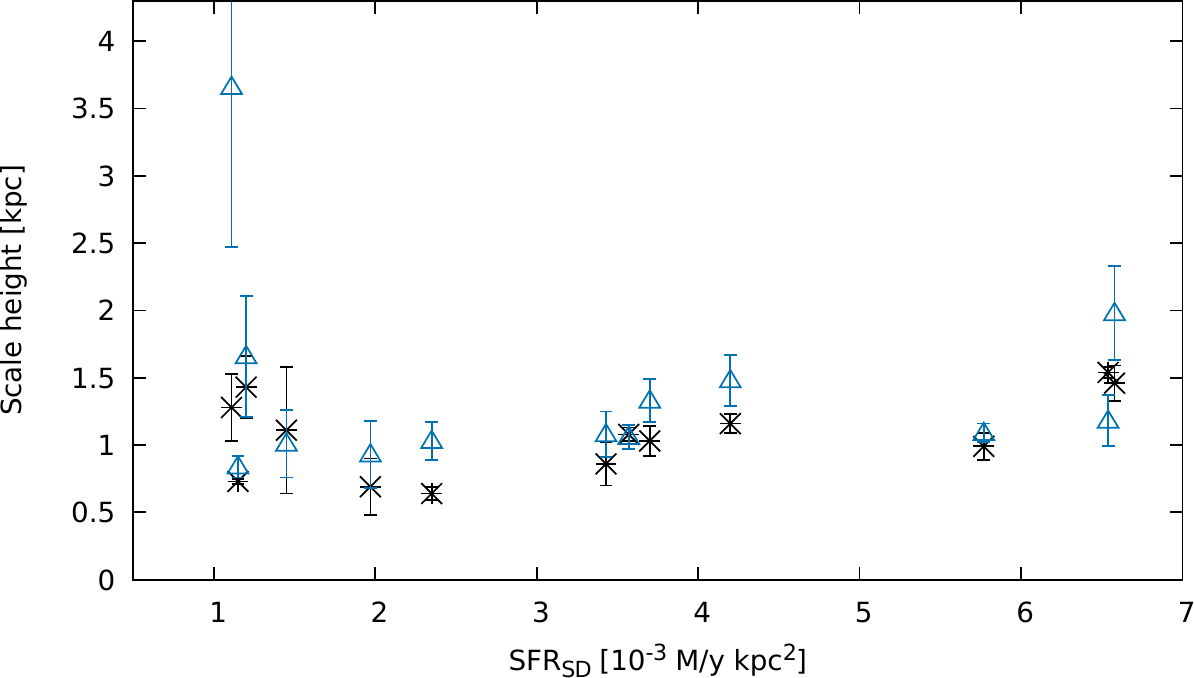}
      \caption{Radio scale heights vs. star formation rate surface density. C-band data are represented by black
      asterisks, L-band data by blue triangles. The L-band outlier is again NGC~3003. No clear correlation is visible.
      }

         \label{SFRSD-h}
   \end{figure}

\begin{figure}
   \centering
   \includegraphics[width=0.95\columnwidth]{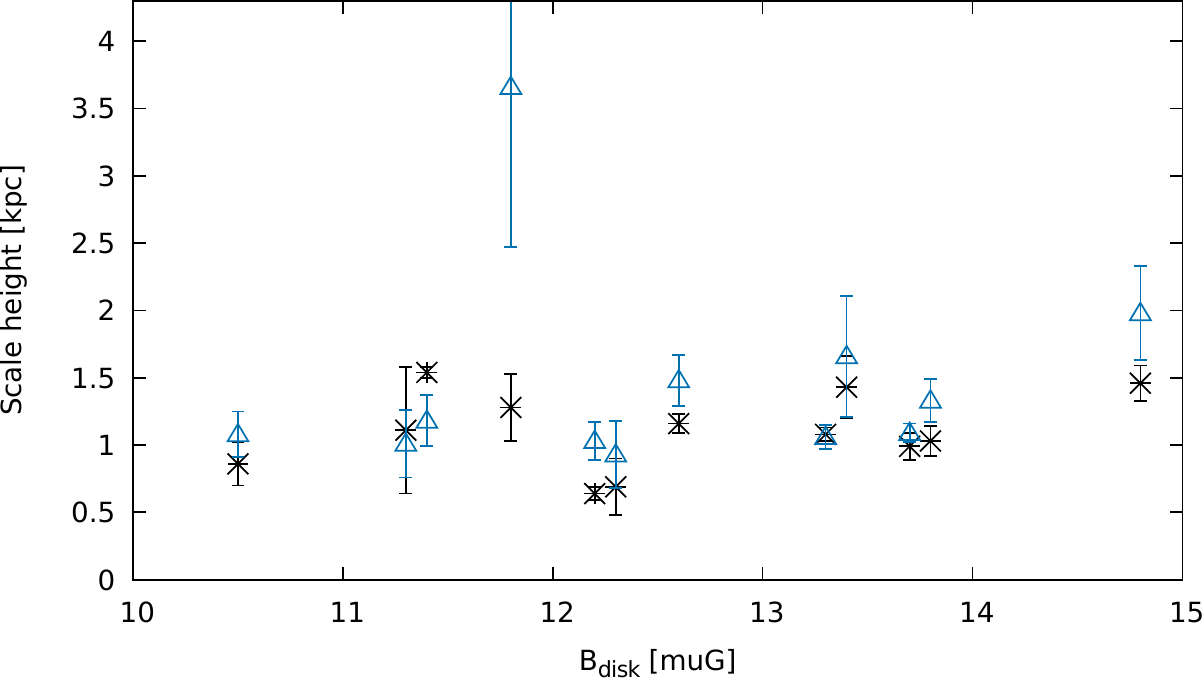}
      \caption{Scale height vs. the total magnetic field strength in the disk B$_\mathrm{t(disk)}$. C-band data are represented by black
      asterisks, L-band data by blue triangles. UGC~10288 was not included in the plot because of its significantly lower magnetic field strength
      B$_\mathrm{disk} = 6.7 \mu$ G.
      }
         \label{Bdisk-h}
   \end{figure}

As star formation in the disk is thought to be the source of the accelerated CRE in the disk and halo, we tested our data sets for possible correlations of the radio
scale heights with the star formation activity in the disk (SFR and SFR$_\mathrm{SD}$), and further with
the total magnetic field strength (B$_\mathrm{t}$, and that in the disk B$_\mathrm{t(disk)}$).
We found no clear dependence of the halo scale heights and SFR$_\mathrm{SD}$ (as shown in Fig.~\ref{SFRSD-h}) or SFR, nor with B$_\mathrm{t(disk)}$ (see Fig.~\ref{Bdisk-h})
or B$_\mathrm{t}$.

\begin{figure}
   \centering
  \includegraphics[width=0.95\columnwidth]{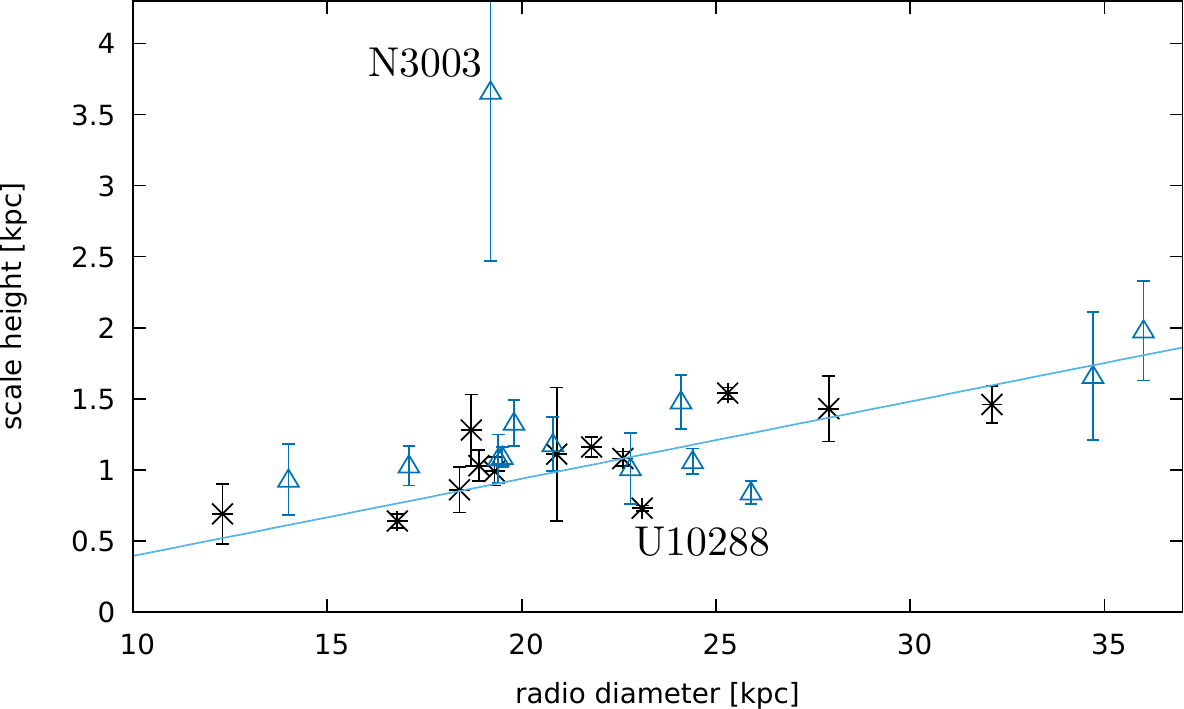}
     \caption{Halo scale height vs. the radio diameter. C-band data are represented by black asterisks,
      L-band data by blue triangles. A weighted linear fit to all data together has a slope of $0.05 \pm 0.01$ with intercept of about zero ($-0.15 \pm 0.29$).
      For clarity, only the errors in scale height are presented; those in diameter are identical with Figure~\ref{dia-length}.
                 }
        \label{dia-height}
  \end{figure}

As discussed in Sect.~\ref{sec:scale length}, we found that both the diameter and the halo scale heights increase with distance. A direct comparison of the scale heights
with the diameters (both measured in kpc)
shows that the scale heights increase linearly with the galaxy's size (radio diameter) (see Figure~\ref{dia-height}). This cannot be explained by any selection effect
due to our
selection criteria, as discussed in Sect.~\ref{sec:scale length}. We fitted the radio data of C-band and L-band together with a weighted linear fit with a slope
of $0.05 \pm 0.01$, an intercept of about zero (reduced $\chi^2 = 4.1$). The outlier is again NGC~3003 in L-band. UGC~10288 has the smallest scale heights in both bands
compared to its large diameter. The radio scale heights also increase with the optical diameter (see Table~\ref{diameter}) in the blue
light, as given by the 25th mag~$\mathrm{arcsec}^2$ \citep{tully1988}, even though less significant. Table~\ref{diameter} already shows that the optical diameters are not
tightly correlated with the radio diameters.

\begin{figure}
   \centering
    \includegraphics[width=0.95\columnwidth]{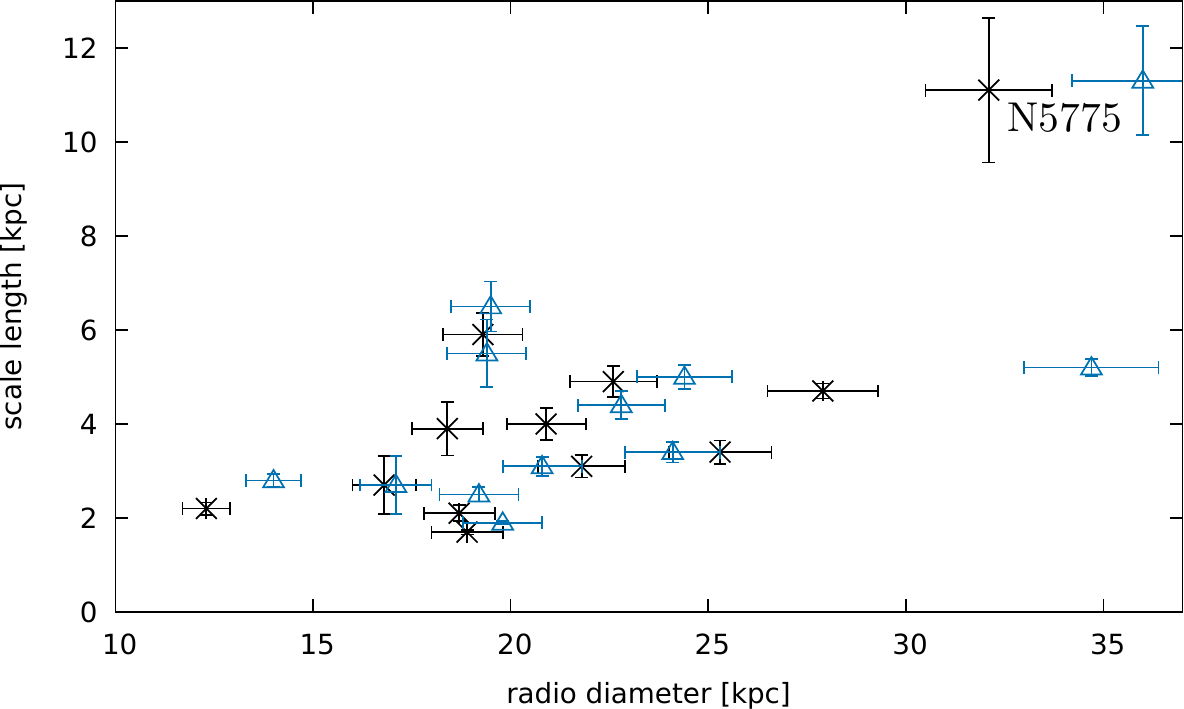}
      \caption{Scale length vs. the radio diameter. C-band data are represented by black asterisks,
      L-band data by blue triangles. A weighted linear fit to all data together has a slope of $0.16 \pm 0.04$ with intercept of about zero ($-0.15 \pm 0.89$).
      UGC~10288 is omitted (see note a of Table~\ref{scalelength}).
              }
         \label{dia-length}
   \end{figure}

A similar trend is visible between the scale lengths and the radio diameters in our sample galaxies (see Figure~\ref{dia-length}), though the spread of the scale
lengths is larger than that in scale height. We consider these relations to be physical.

\begin{figure}
   \centering
   \includegraphics[width=0.9\columnwidth]{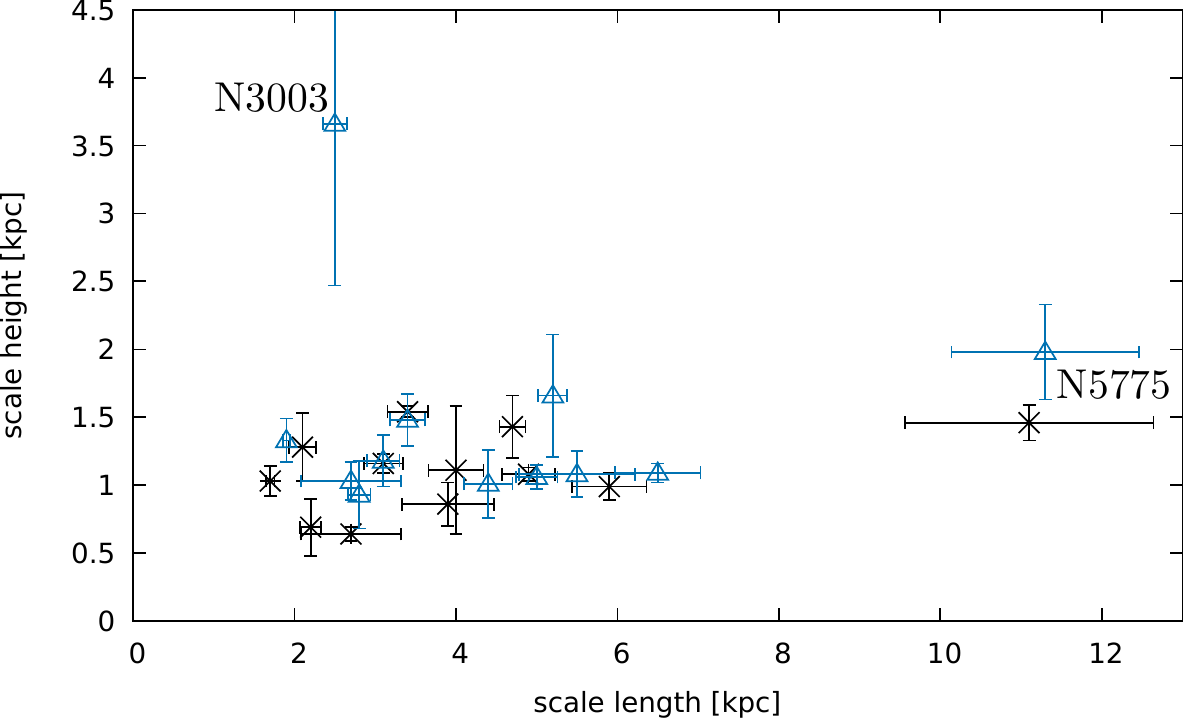}
      \caption{Halo scale lengths vary over a wider range of values than scale heights, especially in C-band. C-band data are represented by black
      asterisks, L-band data by blue triangles.
              }
         \label{length-height}
   \end{figure}

A comparison between the scale heights and scale lengths is presented in Figure~\ref{length-height} for both bands. It also shows a linear correlation.

In order to exclude a bias
due to the size of the galaxy, we determined the ratios between the radio scale heights from Table~\ref{halo scale heights}, for C- and L-band each, and the radio
diameters $d_{\rm r}$ as given in Table~\ref{diameter}. We define this quantity as the normalized scale height $\tilde{h} = 100 \cdot h / d_{\rm r}$
which is presented in Table~\ref{halo scale heights} for C- and L-band together with its error $\Delta \,\tilde{h}$ according to error propagation. We choose the radio diameters
instead
of the optical diameters, as the radio diameters also correct for spectral index effects between the observations at C-band and L-band.

The normalized scale heights $\tilde{h}$ are - different from the scale heights - independent of the diameters of the galaxies. A closer look at the $\tilde{h}$ values
reveals that they are about constant, except for the outliers NGC~3003 at L-band and UGC~10288: the mean value in C-band (L-band) of 5.22 (5.58) has a spread of only 
0.77 (0.80) which is smaller than most of their individual errors $\Delta \tilde{h}$ in Table~\ref{halo scale heights}. Plots of the normalized scale heights against 
the SFR, SFR$_\mathrm{SD}$, B$_\mathrm{t}$, and B$_\mathrm{t(disk)}$ do not show further correlations of the normalized scale heights with one of these parameters.

\begin{figure}
   \centering
   \includegraphics[width=0.9\columnwidth]{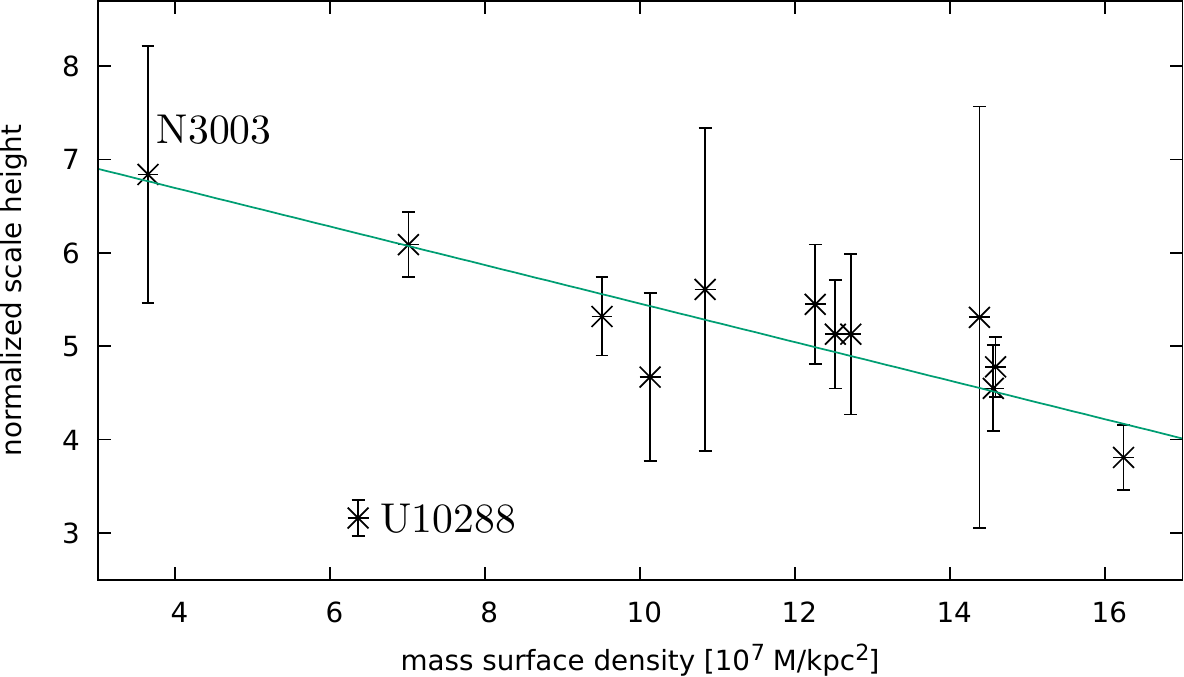}
      \caption{Normalized scale heights in C-band vs. the mass surface density measured in solar masses. The weighted linear fit to the data (excluding UGC~10288) has a
      slope
      of $-0.21 \pm 0.03$ (reduced $\chi^2 = 0.37$).
              }
         \label{MSD-htildeC}
   \end{figure}

\begin{figure}
   \centering
   \includegraphics[width=0.9\columnwidth]{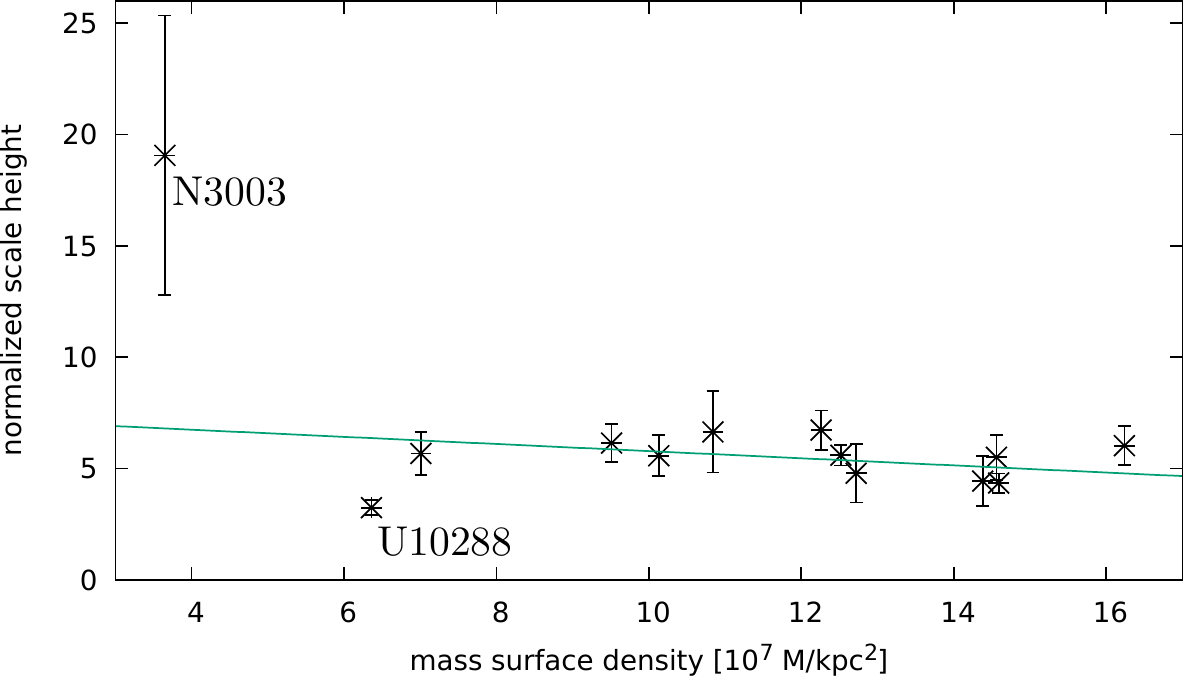}
      \caption{Normalized scale heights in L-band vs. the mass surface density measured in solar masses. The weighted linear fit to the data (excluding UGC~10288 and
      NGC~3003)
      has a slope of $-0.16 \pm 0.10$ (reduced $\chi^2 = 0.96$).
              }
         \label{MSD-htildeL}
   \end{figure}

We also determined the total mass surface density of each galaxy by $MSD = M / \pi (d_{25} / 2) ^2$ where $d_{25}$ is the blue isophotol
diameter as given in Table~\ref{diameter} and $M$ is the total mass \citep{irwin+2012a}. The values for the total mass surface densities vary significantly within the galaxy
sample and are presented in Table~\ref{bfeld}. While the halo scale heights do not correlate with $MSD$, we found an anticorrelation between the normalized scale heights and 
$MSD$ as presented in Figure~\ref{MSD-htildeC} for C-band and in Figure~\ref{MSD-htildeL} for L-band. Remarkably, the fit parameters for C-band and L-band (without the strongly 
deviating L-band value for NGC~3003) are equal within their errors, hence we could have fitted them together but decided to present them separately. UGC~10288 was not included 
into the fit as its flux density was found to be below the flux density limits of the CHANG-ES sample, as discussed in Sect.~\ref{sec:scale length}. The fitted parameters of 
the linear fit to the values of both bands together is $m = -0.19 \pm 0.03$ (slope) and $b = 7.50 \pm 0.44$ (intercept) with a reduced $\chi^2 = 0.72$.

We conclude that the radio scale heights $h$ of the halo depend mainly and linearly on the radio diameter of the galaxy. If this dependence is eliminated by taking the
normalized scale heights $\tilde{h}$ instead, we detected a tight and linear anticorrelation of $\tilde{h}$ with the mass surface density of the galaxy. This will be further 
discussed in Sect.~\ref{sec:discussion}.

Similar to the normalized scale height, we defined the normalized scale length $\tilde{l} = 100 \cdot l / d_{\rm r}$. The values are presented in Table~\ref{scalelength}
for C- and L-band, together with their errors according to error propagation. Their values vary, however, over a wider range. Their spread in C-band (L-band) is 7.5 (7.6), hence
much greater than the individual errors $\Delta \tilde{l}$ in Table~\ref{scalelength}.

\begin{figure}
   \centering
   \includegraphics[width=0.95\columnwidth]{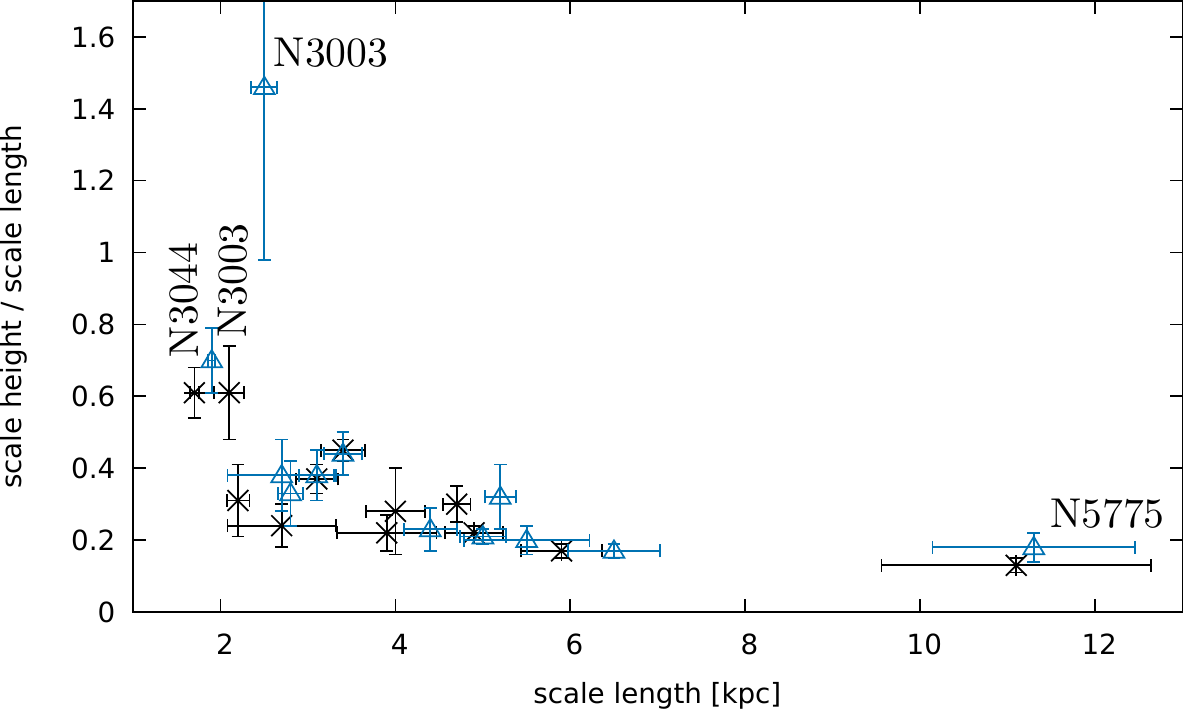}
      \caption{Scale heights in units  of scale lengths vs. scale lengths. C-band data are represented by black
      asterisks, L-band data by blue triangles.
              }
         \label{h/l-length}
   \end{figure}

\begin{figure}
   \centering
   \includegraphics[width=0.95\columnwidth]{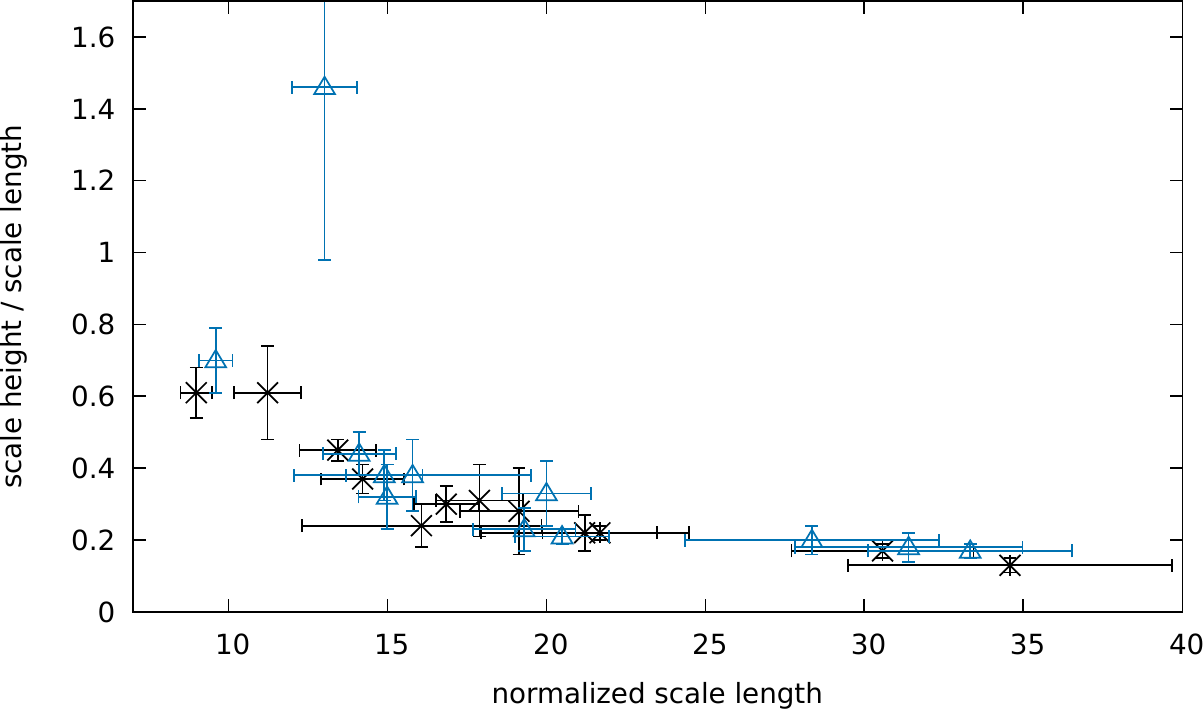}
      \caption{Scale heights in units of scale lengths vs. scale length divided by the radio diameter (i.e. normalized scale length). C-band data are represented by black
      asterisks, L-band data by blue triangles. The outlier is again NGC~3003 in L-band. This plot is free of diameter dependence.
              }
         \label{h/l-normlength}
   \end{figure}

When measuring the radio scale heights in units of scale lengths they systematically decrease with scale lengths (and also with the normalized scale lengths) as shown in 
Figure~\ref{h/l-length}. The same trend is visible
when plotting the scale heights in units of scale length against the normalized scale lengths (Figure~\ref{h/l-normlength}). There, diameter effects are eliminated for
both coordinates. We conclude from these figures that galaxies with smaller scale lengths are more spherical in the radio emission than those with larger scale lengths.
We tested whether this relation also depends on the SFR$_\mathrm{SD}$ or the Hubble type of the galaxies without a positive result.

\begin{figure}
   \centering
   \includegraphics[width=0.95\columnwidth]{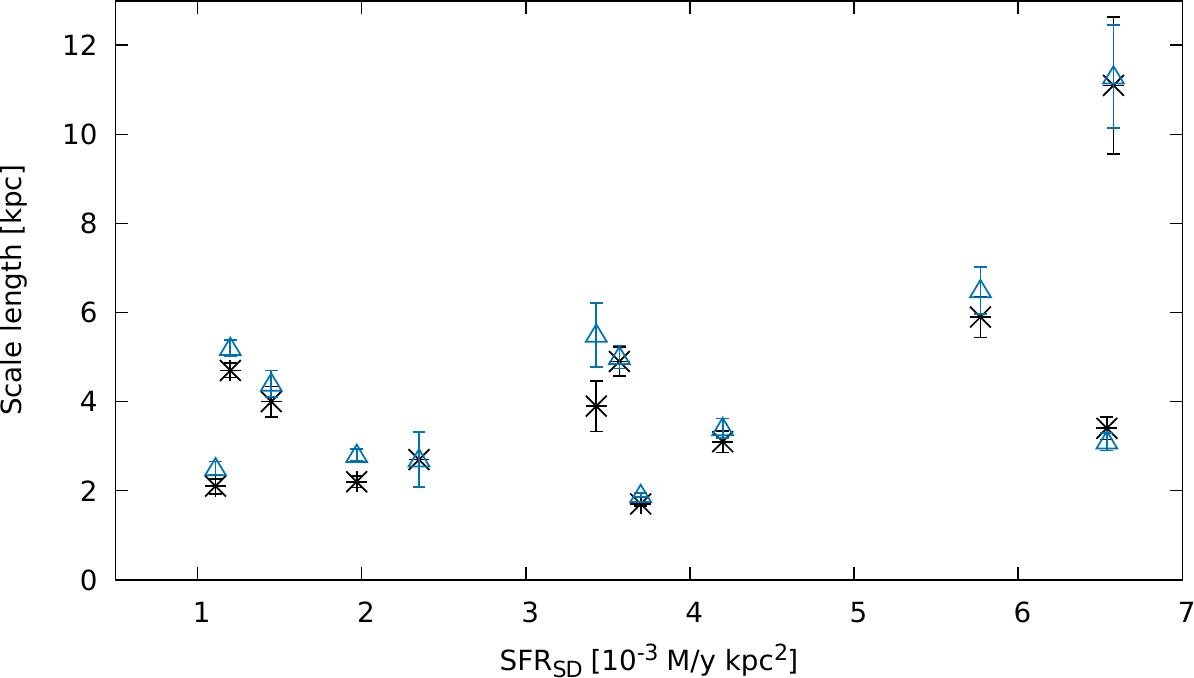}
      \caption{Scale length vs. the star formation rate density  SFR$_\mathrm{SD}$ without UGC~10288. C-band data are represented by black
      asterisks, L-band data by blue triangles.
              }
         \label{SFRSD-l}
   \end{figure}

\begin{figure}
   \centering
   \includegraphics[width=0.95\columnwidth]{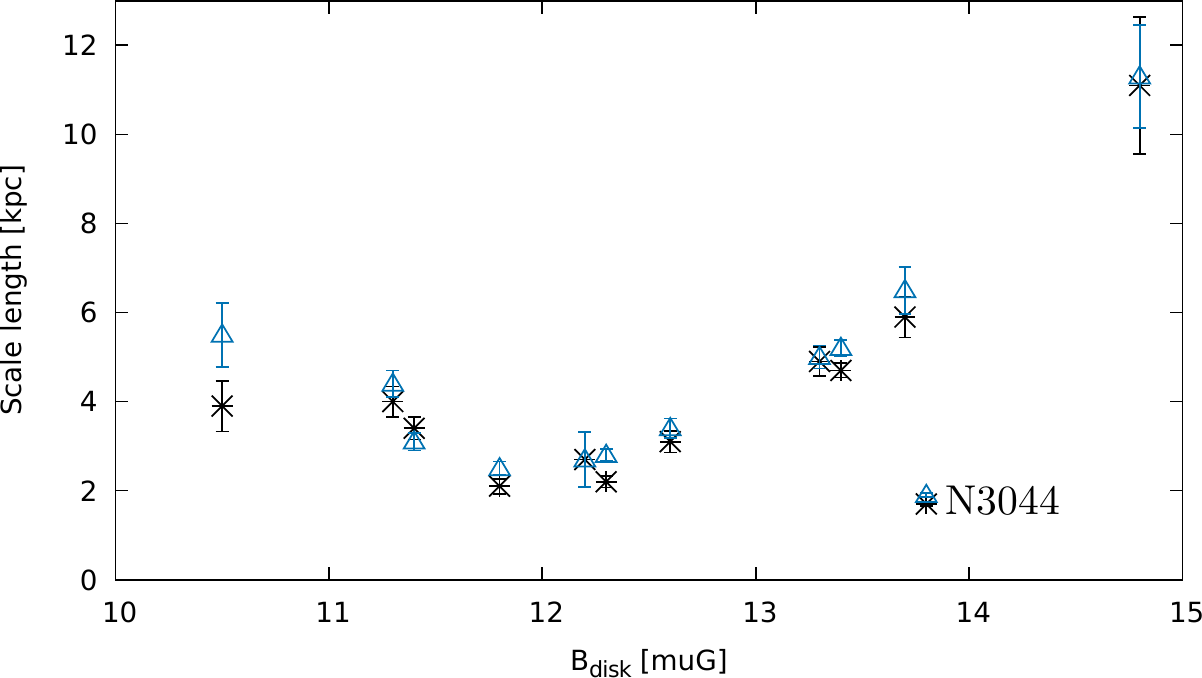}
      \caption{Scale length vs. the total magnetic field strength in the disk B$_\mathrm{t(disk)}$ without UGC~10288. C-band data are represented by black
      asterisks, L-band data by blue triangles.
              }
         \label{Bdisk-l}
   \end{figure}

The question remains: what determines the scale lengths in a galaxy? We tested whether the scale lengths depend on the star formation or magnetic field strength,
similar to the test for the scale heights. The dependence of the scale lengths on SFR$_\mathrm{SD}$ is shown in Figure~\ref{SFRSD-l} and that on
B$_\mathrm{t(disk)}$ in Figure~\ref{Bdisk-l}. While no clear correlation of the scale lengths with SFR, SFR$_\mathrm{SD}$, or B$_\mathrm{t}$ is visible, a linear increase in the
scale lengths with
B$_\mathrm{disk} \gtrsim 11.5 \, \mu G$ is indicated. The outlier N3044 in Figure~\ref{Bdisk-l} shows an outstanding ordered X-shaped magnetic field with strong vertical
components
above and below its central region in our sample \citep{wiegert+2015} which may easily have caused an overestimation of B$_\mathrm{t(disk)}$ by $(1-2)~\mu G$.

\section{Discussion}
\label{sec:discussion}

Within this work we detected for the first time that the radio scale height increases linearly with radio diameter, which is surprising.
We could not find an obvious physical explanation that would not simultaneously imply a correlation of the
halo scale height with SFR or SFR$_\mathrm{SD}$.

In order to exclude this correlation from the further analysis we defined the normalized scale height $\tilde{h}$ and normalized scale length $\tilde{l}$ as
described in Sect.~\ref{sec:correlations}. We found that the normalized scale heights are very similar and that the radio scale height depends mainly on the
radio diameter of the galaxy. Even though our sample may be biased towards brighter galaxies due to the selection criteria of the CHANG-ES sample (as discussed in
Sect.~\ref{sec:scale length}), our sample contains galaxies with low SFR and SFR$_\mathrm{SD}$, also at large distances. We found, remarkably, that neither the scale
height $h$ nor $\tilde{h}$ depend on SFR or SFR$_\mathrm{SD}$. We conclude that the radio halo properties are not simply dictated by the star formation activity.

However, we found a tight and linear anticorrelation of the normalized scale height $\tilde{h}$ with the mass surface density $MSD$ of the galaxies. The mass surface
density is a measure of the gravitational potential of the galaxy with the baryonic mass being mainly concentrated in the disk of a galaxy. A similar decrease in 
$\tilde{h}$ with an increase of $MSD$ at both frequency bands as shown in Figures ~\ref{MSD-htildeC} and \ref{MSD-htildeL} can be regarded as an observational indication 
for a gravitational deceleration of the CRE outflow on its way up into the halo.

The radio-weak galaxy UGC~10288 with its radio intensity well below the CHANG-ES flux density limit has a significantly lower magnetic field strength than
the sample average (see Sect.~\ref{sec:field strength}). Though it does not deviate from the sample with respect to its value for the halo scale height or to its
SFR or SFR$_\mathrm{SD}$, it has the smallest value for the normalized scale height $\tilde{h}$.
The reason for this surprising result could be an observational bias of our sample towards
radio-bright galaxies. UGC~10288 may represent a class of radio-weak galaxies that still have a significant level of star formation.
Although there is general consensus that star formation drives the amplification of turbulent magnetic field (called the small-scale dynamo), other still unexplored factors
may affect the efficiency of field amplification.

From the frequency dependence as discussed in Sect.~\ref{sec:frequency} we conclude that the clearest candidates for an escape-dominated halo are NGC~4157,
NGC~4217, and NGC~4302. Their scale heights are similar at both frequencies. In this case, the CRE leave the halo so fast that no significant frequency
dependence due to synchrotron losses can be observed. This means that the adiabatic loss time of CRE is smaller than their
synchrotron life time within the observed distances from the galactic plane. Using Eq.\,\ref{tsyn}, the synchrotron life time of CRE is
$t_{\rm syn} \simeq 1.39 \, 10^7\,$yr for C-band and $B = 10\, \mu$G. In escape-dominated halos, we can assume equipartition between magnetic fields and cosmic rays,
so that the scale height of the cosmic-ray electrons $h_{\rm CRE}$ is about twice the observed synchrotron scale height $h_z$, which is about 1~kpc for the three galaxies 
mentioned above (see Table~\ref{halo scale heights}). Hence, we can estimate a mean cosmic-ray bulk speed $\mathrm{v_{CRE}}$ (corresponding to the velocity of a galactic wind) 
of $\mathrm{v_{CRE}} = 2 h_z / t_{\rm esc} > 2 h_z / t_{\rm syn} \simeq 140\,$km/s.

The adiabatic loss time depends only on the reciprocal velocity gradient: $t_{\rm ad} = 3 \left(\frac{\partial {\mathrm{v}}}{\partial z}\right)^{-1}$ \citep{heesen+2009}.
Assuming a constant increase in the CRE velocity with height gives $\mathrm{v_{\rm wind}} = 6 h_z / t_{\rm ad} = 420\,$km/s, significantly higher than 140~km/s
and above the escape velocity from the disk, which is about 280~km/s for a rotation speed of 200~km/s \citep{heesen+2009}.

On the other hand, if the scale heights mainly depend on the diameters of the galaxy, we should multiply the L-C-band scale height ratio of Table~\ref{halo scale heights}
by the inverse of the L- to C-band diameter ratio of Table~\ref{diameter} before performing the frequency analysis. This correction for the diameters leads, however, to
about equal scale heights in C-band and L-band,
as indicated already in Figure~\ref{dia-height} by the fact that the values in C- and L-band can be fitted  together well by one linear fit. This directly leads to the
conclusion that all galaxies have an escape-dominated halo with convective CRE propagation (see Sect.~\ref{sec:frequency}).

As discussed in Sect.~\ref{sec:frequency} the scale height $h_{\rm z}$ of a synchrotron energy loss-dominated halo depends not only  on the observing frequency, but also on
the strength of the total magnetic field. According to Eq.\,\ref{diffusion.energydep}, Eq.\,\ref{diffusion.notenergydep}, and Eq.\,\ref{convection}, $h_{\rm z}$ would
decrease with increasing magnetic field strength for a synchrotron energy loss-dominated halo. This is not visible in Figure~\ref{Bdisk-h}. The CRE in the halos of our sample
galaxies are not dominated by synchrotron losses, consistent with the lack of frequency dependence as discussed above.

The magnetic field strength may be another important parameter for the evolution of a radio halo. The galaxy with the  smallest magnetic field strength in
the sample by far is UGC~10288. It is the galaxy whose values lie significantly beyond the fitted curve in Figure~\ref{dia-height} and hence has the smallest normalized scale
height within the sample (see Table~\ref{halo scale heights}) significantly outside the mean normalized scale height of $5.22 \pm 0.77$. We recall here that UGC~10288
deviates only in its magnetic field strength (and radio intensity) from the other sample galaxies, but not in its radio scale height nor its star formation activity.

Furthermore, NGC~3044 can be regarded as an outlier in Figure~\ref{Bdisk-l}. Its scale length is relatively small for its large $B_\mathrm{disk}$. In comparison with the
other sample galaxies, NGC~3044 shows strongly polarized intensity indicating a well-ordered X-shaped magnetic field with strong vertical components above and below its
central region \citep{wiegert+2015}. Such a field configuration should help the CRs to escape from the disk into the halo. Even here, the radio scale height is
comparable to the other sample galaxies, but the scale length is smaller. With this single galaxy we cannot clarify whether this is just an accidental coincidence or
if it indicates a causal influence on the development on the radio scale lengths. However, this finding fits into the given interpretation of Figure~\ref{SFRSD-l} and
Figure~\ref{Bdisk-l} in Sect.~\ref{sec:correlations} that the scale lengths grow with increasing magnetic field strength in the disk and SFR$_\mathrm{SD}$ unless
the field configuration supports the CR escape directly by strongly ordered, vertical magnetic field components.

This suggests that especially the magnetic field strength in the disk B$_\mathrm{t(disk)}$ and the field structure of an ordered or large-scale magnetic field in the halo are
important parameters for the evolution of a radio halo, e.g. by supporting the CR outflow or increasing the wind velocity.

For the estimates of the total magnetic field strengths we assumed energy equipartition between the CRE and the magnetic field in the halo. This is not necessarily a
given. A faster
decrease in the magnetic field strength than expected by equipartition would also decrease the observed radio emission. This possibility has been included
and can be modelled by the `SPINNAKER' code of \cite{heesen+2016} which also needs spectral indices as inputs. On the other hand, strong synchrotron losses increase
the ratio $K$ of cosmic-ray protons and electrons. Our assumption of the standard value of $\rm {K}=100$ would then underestimate the field strength in the halo.
The analysis of the CHANG-ES galaxies with this method is planned and will be published in a forthcoming paper.

\section{Summary and conclusions}
\label{sec:summary}

Within this work we defined a subsample of the 35 CHANG-ES galaxies as being smaller than $4.5 \arcmin$ in diameter with an inclination larger than $80 \degr$
with extended disk emission not severely disturbed or dominated by nuclear activity and/or interaction. This leads to a sample of 13 galaxies in total.
For these galaxies we determined the vertical scale heights for the thin disk and halo from C-band and L-band observations with an angular resolution of about
$10 \arcsec$ in a consistent way with a more sophisticated method than used before. Different from the halo scale heights, the relative errors of the values for
the thin disk are too high to be discussed further.

We checked our results for possible systematic effects of missing large-scale flux density which could play a role mainly at C-band, but we found no indication of a
significant effect. We determined the scale height ratios of the results at both frequency bands and discussed the frequency dependence of the halo
scale heights with respect to the CR transport and energy loss mechanisms.

Assuming energy equipartition between the magnetic field and cosmic rays we determined the total magnetic field strengths B$_\mathrm{t}$ for the whole galaxy
and for the disk alone B$_\mathrm{t(disk)}$ for all sample galaxies. Finally, we tested our data sets for possible correlations.

Our main results are as follows:

\begin{itemize}
 \item The averaged values for the radio scale heights of the halo are $1.1\, \pm \, 0.3$~kpc in C-band and $1.4\, \pm \, 0.7$~kpc in L-band.
 \item The halo scale height and scale length increase linearly with the radio diameters.
 \item In order to correct for a bias due to the radio size of the galaxy, we defined a normalized scale height $\tilde{h} = 100 \cdot h / d_{\rm r}$ and a
 normalized scale length $\tilde{l} = 100 \cdot l / d_{\rm r}$.
 The normalized scale heights $\tilde{h}$ are equal within the scale height errors for our sample galaxies; only UGC~10288 has a slightly lower value.
 \item Those spiral galaxies in our sample with smaller scale lengths are more spherical in the radio emission, while those with larger scale heights are flatter.
 \item The total magnetic field strengths B$_\mathrm{t}$ (B$_\mathrm{t(disk)}$) are in the small range of $9-11~\mu$G ($11-15~\mu$G). Only UGC~10288 has a
 weaker field of about $6~\mu$G.
 \\
\end{itemize}

From the analysis of the frequency dependence, the clearest candidates for an escape-dominated halo are NGC~4157, NGC~4217, and
NGC~4302, for which the wind velocities are estimated using the adiabatic loss time to be up to about 420~km/s at the height of 1~kpc (the halo scale height).
This value seems to be rather high and may reduce to up to 1/3 of this value if adiabatic losses occur not only in z-direction, but are also significant in all three dimensions.
In this case the velocity would be equal to 140~km/s which is the velocity required for a synchrotron energy loss-dominated halo.
We conclude that the adiabatic losses are mainly in the z-direction and the wind speeds are $\gg 140$~km/s,
up to 420~km/s, hence they may exceed the typical escape velocity of 300~km/s.

The scale height ratios between L-band and C-band range mainly between 1 and 2. This range is expected for synchrotron energy loss-dominated or escape-dominated halos.
However, if the scale height ratios are corrected for the diameter dependence, they reduce to unity. This is expected for escape-dominated halos with convective propagation.
As  the scale height--magnetic field dependence also does not fit to synchrotron energy loss-dominated halos, we conclude that the sample galaxies (except possibly UGC~10288)
have an escape-dominated halo with convective propagation.\\ 

In summary we make the following conclusions: 
\begin{itemize}
 \item The halo scale height of our sample galaxies increases mainly with the radio diameter of the galaxies. The physical reason for this remarkable result remains unexplained.
 \item When corrected for the diameter effect, the radio scale height shows a tight and linear anticorrelation with the mass surface density,
 indicating a gravitational deceleration of the CRE outflow from the galactic disk, e.g. a lowering of the wind velocity.
 \item All sample galaxies are consistent with an escape-dominated halo with convective propagation.
 \item A higher SFR or SFR$_{SD}$ does not lead directly to a faster CR transport (i.e. larger wind velocity).
\end{itemize}

In contrast to the scale heights, neither the scale lengths $l$ nor the normalized scale length $\tilde{l}$ are constant within our sample galaxies. We found that $\tilde{l}$
does not depend on SFR, SFR$_\mathrm{SD}$, or B$_\mathrm{t}$. A linear increase in the scale lengths with the magnetic field strength in the disk for
B$_\mathrm{t(disk)} \gtrsim 11.5 \mu G$ is indicated.

\begin{acknowledgements}

We thank Andrew Fletcher and Aritra Basu for fruitful discussions, and Aritra Basu also for helpful comments on the manuscript. We acknowledge a helpful suggestion of the
anonymous referee. This research has made use of the NASA/IPAC Extragalactic Database (NED) which is operated by the Jet Propulsion Laboratory, California Institute of Technology, 
under contract with the National Aeronautics and Space Administration. RAMW acknowledges support from the National Science Foundation through grant AST-1615594.

\end{acknowledgements}

\bibliographystyle{aa}
\footnotesize

\end{document}